\documentclass[10pt, preprint, twocolumn]{emulateapj}		
\usepackage{color}

\usepackage{style_wei_symbols}
\usepackage{style_wei}

\usepackage{natbib}
\begin{document}


\title{Quasi-periodic Fast-mode Wave Trains Within a Global EUV Wave 
and Sequential Transverse Oscillations Detected by {\it SDO}/AIA}








\shorttitle{Quasi-periodic Wave Trains Within A Global EUV Wave Traveling Through A Coronal Cavity}
\shortauthors{Liu et al.}
\slugcomment{Accepted by ApJ, April 24, 2012}

\author{Wei Liu\altaffilmark{1}\altaffilmark{2}, Leon Ofman\altaffilmark{3}\altaffilmark{4}, 
        Nariaki V.~Nitta\altaffilmark{1}, Markus J.~Aschwanden\altaffilmark{1},
        Carolus J.~Schrijver\altaffilmark{1},
        Alan M.~Title\altaffilmark{1}, and Theodore D.~Tarbell\altaffilmark{1}}

\altaffiltext{1}{Lockheed Martin Solar and Astrophysics Laboratory, 
  3251 Hanover Street, Palo Alto, CA 94304, USA; weiliu@lmsal.com}	
\altaffiltext{2}{W.~W.~Hansen Experimental Physics Laboratory, Stanford University, Stanford, CA 94305, USA}
\altaffiltext{3}{Catholic University of America and NASA Goddard Space Flight Center, 
  Code 671, Greenbelt, MD 20771, USA}
\altaffiltext{4}{Visiting Associate Professor, Department of Geophysics and Planetary Sciences, Tel
Aviv University, Tel Aviv 69978, Israel}


\begin{abstract}	
	

We present the first unambiguous detection of quasi-periodic wave trains
within the broad pulse of a global EUV wave (so-called ``EIT wave") occurring on the limb.
These wave trains, running {\it ahead of} the lateral CME front of 2--4 times slower,
coherently travel to distances $\gtrsim$$R_\sun/2$ along the solar surface,
with initial velocities up to $1400 \kmps$ decelerating to $\sim$$650 \kmps$.	
The rapid expansion of the CME initiated at an elevated height of 110~Mm 	
produces a strong downward and lateral compression,		
which may play an important role in driving the primary EUV wave and shaping its front
forwardly inclined toward the solar surface. 
The waves have a dominant 2~min periodicity that matches the X-ray flare pulsations,
suggesting a causal connection.	
The arrival of the leading EUV wave front at increasing distances produces an uninterrupted
chain sequence of deflections and/or transverse (likely fast kink mode) oscillations of local structures,
including a flux-rope coronal cavity and its embedded filament
with delayed onsets		
consistent with the wave travel time at an elevated (by $\sim$50\%) velocity within it.
This suggests that the EUV wave penetrates through a topological separatrix surface into the cavity, 
unexpected from CME caused magnetic reconfiguration.	
These observations, when taken together, provide compelling evidence of the fast-mode MHD wave nature of
the {\it primary (outer) fast component} of a global EUV wave, running ahead of the {\it secondary (inner) slow}
component of CME-caused restructuring.

\end{abstract}


\section{Introduction}
\label{sect_intro}

%



Waves or traveling disturbances on {\it global} scales in the magnetized solar atmosphere
have been a subject of intensive study over the past half century, partly because of their possible roles
in transporting energy	
and diagnostic potential in inferring physical conditions on the Sun
via coronal seismology \citep{			
Uchida.coronal-seismology.1970PASJ...22..341U, Roberts.coronal-seismology.1984ApJ...279..857R,
AschwandenM2004psci.book.....A, Nakariakov.wave-review.2005LRSP....2....3N}.

\subsection{Global EUV (``EIT") Waves}
\label{sect_intro:eitwv}

One of the most spectacular traveling disturbances in the corona is expanding 	
extreme ultraviolet (EUV) enhancements across a large fraction of the solar disk,
discovered by the \soho EUV Imaging Telescope (EIT) 
and often called ``EIT waves"
\citep{Moses.EIT-wave.1997SoPh..175..571M, 		
ThompsonB.EIT-wave-discover.1998GeoRL..25.2465T}
or {\it global EUV waves}, which we adopt in this paper (or ``EUV waves" for short).  
Their counterparts were also imaged at other wavelengths, including
soft X-rays \citep{KhanAurass.SXT-wave.2002A&A...383.1018K, Narukage.SXT-EIT-wave.2002ApJ...572L.109N} 
and radio emission \citep{WarmuthA.multiw-EIT-wave.2004A&A...418.1101W, 
White.Thompson.Nobeyama-EIT-wave.2005ApJ...620L..63W}.		
Traveling disturbances in the chromosphere
were detected in \Ha \citep{Moreton.wave.1960AJ.....65U.494M, Uchida.Moreton-wave-sweeping-skirt.1968SoPh....4...30U,
Balas.Moreton.wave.2010ApJ...723..587B}
and \ion{He}{1} 10803~\AA\ lines \citep{Vrsnak.HeI.wave.2002A&A...394..299V, Gilbert.HeI.wave.2004ApJ...610..572G}.

Global coronal seismology using large-scale EUV waves
\citep{Ballai.global-corona-seism.2007SoPh..246..177B},
compared with local seismology using loop oscillations, has been little explored,
because of their poorly understood origin and nature. 
Proposed mechanisms include 	
 {\it fast-mode magnetohydrodynamic (MHD) waves} as the coronal counterpart of chromospheric Moreton waves 
\citep{ThompsonB.EIT-Moreton-wave.1999ApJ...517L.151T,
Wills-DaveyThompson.TRACE-EUV-wave.1999SoPh..190..467W, 
WangYM.EIT-fastMHDwave.2000ApJ...543L..89W,	
Wu.EIT-fast-MHD-wave.2001JGR...10625089W, WarmuthA.EIT-Moreton-wave-fast-mode.2001ApJ...560L.105W, 
OfmanThompson.EIT-wave-fast-mode.2002ApJ...574..440O,
GrechnevV.shock-EUV-wave-III.2011SoPh..273..461G, GrechnevV.shock-EUV-wave-I.2011SoPh..273..433G},
and coronal mass ejection (CME) produced {\it current shells or successive 	
restructuring of field lines} \citep{Delannee.EIT-pseudo.wave.2000ApJ...545..512D, 
Delannee.EIT-wave-current-shell.2008SoPh..247..123D,
ChenFP.EIT-wave-MHD.2002ApJ...572L..99C, ChenFP.EIT-wave.2005ApJ...622.1202C, ChenF.Ding.ChenPF.EIS.EITwv.nonWave.2011ApJ...740..116C,
Attrill.EIT-wave-CME-Footprint.2007ApJ...656L.101A, DaiY.EUVI-wave.non-wave.2010ApJ...708..913D}.
Fast modes, for example, as the only MHD waves that can propagate perpendicular to magnetic fields, 
can readily explain the semi-circular shape of global EUV waves
and their observed reflection from coronal holes
and refraction around active regions \citep{ThompsonB.EIT-wave-catalog.2009ApJS..183..225T,
Gopalswamy.EIT-wave-reflect-CH-illusion.2009ApJ...691L.123G, Schmidt.Ofman.3D-MHD-eitwv.2010ApJ...713.1008S,
LiTing.2010Jun07.AIAwave.reflect.2012ApJ...746...13L, OlmedoO.2011-2-15_X2.AIA.EUV.wave.ApJ}.
Details of relevant observations and models can be found in recent reviews
\citep	
{WarmuthA.EIT-wave-review.2007LNP...725..107W, 
WarmuthA.EITwv.short.review.2011PPCF...53l4023W,
VrsnakCliver.corona-shock-review.2008SoPh..253..215V, Wills-Davey.EIT-wave-review.2009SSRv..149..325W, 
GallagherP.LongD.EIT.wave.review.2011SSRv..158..365G, ZhukovA.EIT.wave.review.STEREO.2011JASTP..73.1096Z,
ChenPF.CME.review.2011LRSP....8....1C, Patsourakos.Vourlidas.EIT-wave-review.2012arXiv1203.1135P}.

Early observations of global EUV waves were limited by instrumental capacities,
particularly the low cadence of EIT (12--18~min) and 
small field of view (FOV) of \traceA.
These constraints were alleviated by
the \stereoA/EUV Imagers \citep[EUVI;][]{WuelserJ.STEREO-EUVI.2004SPIE.5171..111W}		
from two vantage points at improved cadence of 150~s, occasionally 75~s
\citep[e.g.,][]{Veronig.STEREO-EUVI-wave.2008ApJ...681L.113V, Veronig.dome-wave.2010ApJ...716L..57V, 
Patsourakos.EUVI-fast-mode-wave.2009ApJ...700L.182P, 	
MaSL.EUVI-wave.2009ApJ...707..503M, 	
TemmerM.3D.stereo.EUVwv.2011SoPh..273..421T, MuhrN.STEREO.EUVwave.2011ApJ...739...89M, 
LongD.STEREO.EUV-wave.Dispersion.2011A&A...531A..42L, 
ZhaoXH.WuST.EIT-wave-2010-01-17.2011ApJ...742..131Z, LiTing.EUVwave.AIA.STEREO.2012RAA....12..104L}.
The more recent {\it Solar Dynamics Observatory} \citep[\sdoA,][]{PesnellD.SDO.mission.2011SoPh..tmp..367P} 
Atmospheric Imaging Assembly \citep[AIA;][]{LemenJ.AIA.instrum.2011SolPhy}
observes the corona at seven EUV wavelengths, 
covering a wide range of temperatures, at high cadence of 12~s and resolution of $1 \farcs5$.
These capabilities have allowed us to study the kinematics and thermal structures of EUV waves
in unprecedented detail \citep[e.g.][]
{LiuW.AIA-1st-EITwave.2010ApJ...723L..53L, ChenFP.Wu.Coexis-fast-mode.2011ApJ...732L..20C,
Schrijver.Aulanier.2011Feb15.X2.AIAwv.2011ApJ...738..167S}		
and continue to shape a better understanding	
of this phenomenon, as we will see in the example presented here.


\subsection{Introduction to This Study}
\label{sect_intro:thisStudy}

\begin{figure}[thbp]      
 \epsscale{1.}
 \plotone{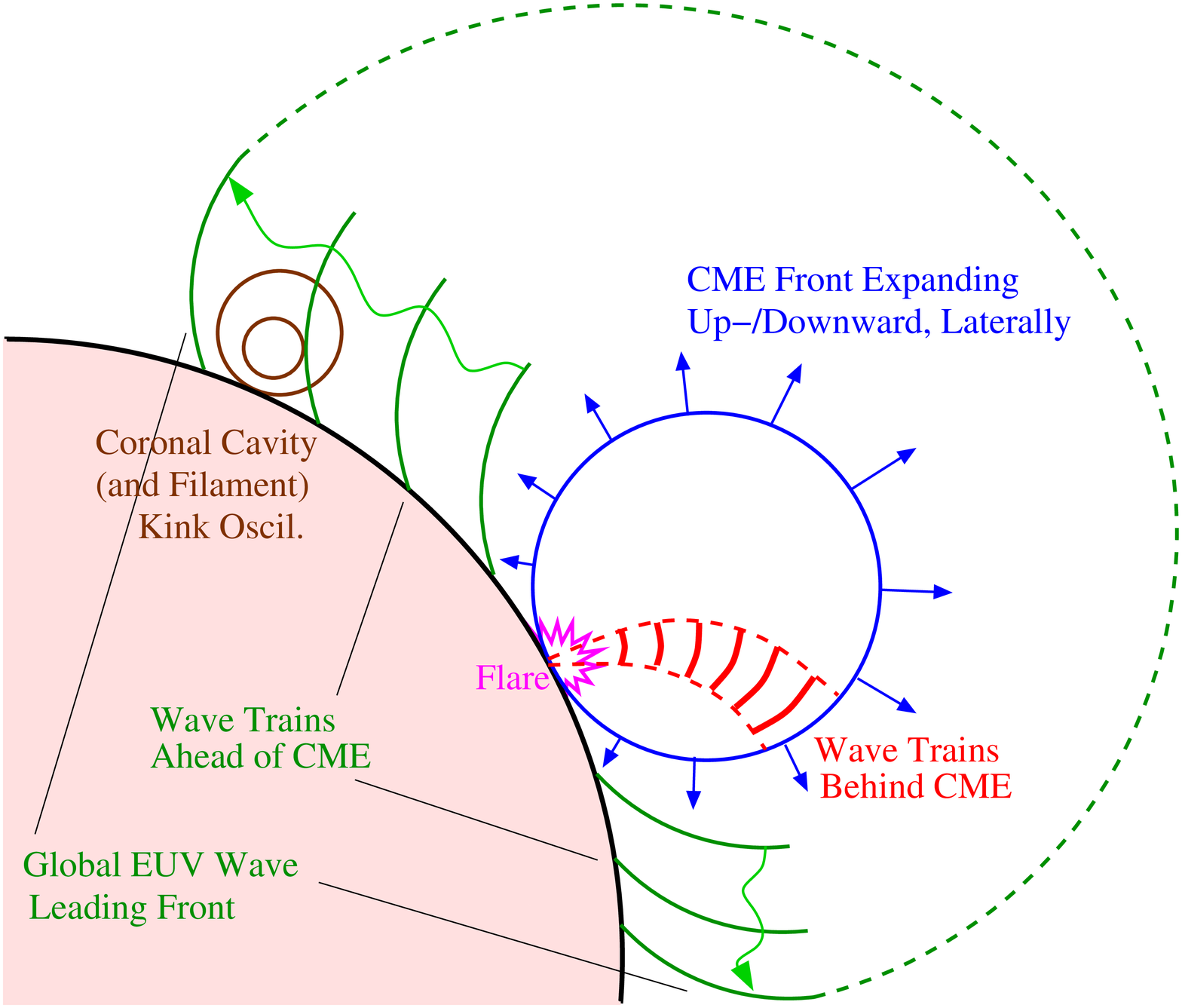}		
 \caption[]{
 Schematic synopsis of the 2010 September 8--9 global EUV wave event,		
 showing quasi-periodic wave trains
 running ahead of (green) and behind (red) the CME front (blue). The former is part of
 the broader EUV wave pulse that travels through a flux-rope coronal cavity,
 instigating its kink-mode oscillations. The CME is initiated at an elevated height
 and produces a downward and lateral compression, driving the low-corona EUV wave.
 The wave trains behind the CME originate at the flare kernel, propagate
 along a narrow funnel, and terminate at the CME front. 
 } \label{cartoon.eps}
 \end{figure}

Observations to date have generally shown a {\it single}, diffuse front in primary EUV waves
\citep[e.g.][]{Wills-Davey.EIT-wave-soliton.2007ApJ...664..556W}.
We report here the first unambiguous detection of {\it multiple}, quasi-periodic wave trains 
within a broad EUV wave pulse, as shown in the schematic of \fig{cartoon.eps}.
These wave trains, comprising fronts forwardly inclined toward the solar surface, 
are repeatedly launched ahead of the lateral CME front
with a 2~min periodicity and travel at velocities up to $\sim$$1400 \kmps$
(see \fig{tslice_QFP.eps}, \sect{sect_QFP:global}).
We will show that the initial CME expansion produces a strong downward compression,
previously unnoticed, which may play an important role 
in driving the EUV wave (see \sect{sect_decouple:wave-generation}).



Filament oscillations 
\citep{Okamoto.EIT-Moreton-wave-filament-oscil.2004ApJ...608.1124O,		
HershawJ.promin.oscil.2011A&A...531A..53H, Tripathi.Isobe.promin.oscil.review.2009SSRv..149..283T}
and loop deflections \citep{Wills-DaveyThompson.TRACE-EUV-wave.1999SoPh..190..467W, 
Patsourakos.EUVI-fast-mode-wave.2009ApJ...700L.182P, 
LiuW.AIA-1st-EITwave.2010ApJ...723L..53L, Aschwanden.Schrijver.AIA.loop.oscil.2011ApJ...736..102A} 		
triggered by global EUV waves or Moreton waves
have been observed at lower cadence or less favorable viewing angles (e.g., on the disk).
AIA's {\it high cadence} allowed us to see, 
for the first time, an uninterrupted chain sequence of deflections and/or oscillations	
of various coronal structures {\it above the limb},
at increasing distances on the path of the global EUV wave. 
In particular, the wave travels across a topological separatrix surface into a coronal cavity 
hosting a filament at an elevated velocity and instigates its bodily oscillations	
(see \fig{tslice_track_oscil.eps}, \sect{sect_oscil}).		

These new observations, when taken together, point to the true wave nature of the
primary EUV wave.
%
We organize this paper as follows.
After an observational overview of the event under study in \sect{sect_overview},	
we focus in \sect{sect_decouple} on the global EUV wave
containing quasi-periodic wave trains running ahead of the CME,
followed by \sect{sect_oscil} on successive structural oscillations on the wave path.
In \sect{sect_QFP}, we compare these wave trains with those in a coronal		
funnel behind the CME \citep{LiuW.FastWave.2011ApJ...736L..13L} and flare pulsations.
We conclude this paper with \sect{sect_conclude}, followed by three Appendixes
for our analysis techniques, the kinematics and temperature and height dependence 
of the EUV wave, and the early expansion of the CME and its role in wave generation.
%

\section{Overview of Observations}
\label{sect_overview}

 \begin{figure}[thbp]      
 \epsscale{0.9}
 \plotone{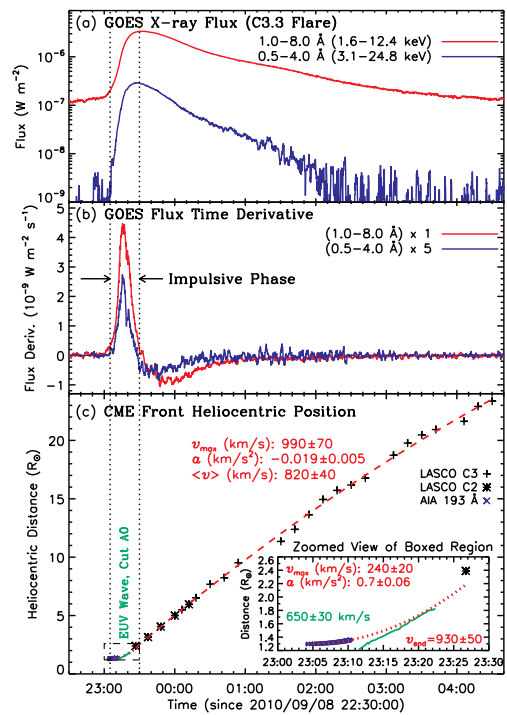}		
 \caption[]{	
  (a) \goes soft X-ray fluxes 	
  and (b) their time derivatives (smoothed with a 2~min box car) for the C3.3 flare.
  (c) Heliocentric positions of the CME front detected by \sohoA/LASCO C3 (plus signs) and C2 (asterisks)
 and AIA (blue crosses, reproduced from \fig{tslice_radial.eps}(h)). 
 The inset is an enlarged view of the boxed region during the
 flare impulsive phase when the CME acceleration and global EUV wave take place.
 Red broken lines are parabolic fits to the LASCO and AIA data labeled with fitted 
 accelerations $a$ and maximum velocities $v_{\rm max}$. 	
 The green curve is the horizontal position of
 the northward global EUV wave front in the low corona, taken from 
 \fig{multi_vfit.eps}(a), 	
 which travels faster than the early CME.	
 } \label{flare_lc.eps}
 \end{figure}
\begin{table}[bthp]	
\scriptsize		
\caption{Event Timeline (2010 September 8--9)}		
\tabcolsep 0.04in	
\begin{tabular}{ll}
\tableline \tableline
 22:40			&	AR core starts gradual bulging	\\
 23:05			&	Onset of rapid CME expansion	\\
 23:05--23:30	&	Flare impulsive phase			\\
 23:10--23:22	&	Flare X-ray pulsations (2~min period)	\\
 23:11/23:14	&	Global EUV wave generated at 110~Mm to \\	
 				&	the North/South of the eruption center	\\
 23:11--23:23	&	Quasi-periodic wave trains launched ahead of CME  \\
 /23:14--23:26	&	at 110~Mm to the North/South of the eruption center	\\
 23:11--23:29	&	QFP wave trains launched from flare kernel behind CME \\
 23:15--02:00+	&	Coronal cavity/filament oscillations	\\
\tableline  \end{tabular}

\label{table_timeline} \end{table} 
 \begin{figure*}[thbp]      
 \epsscale{1.}
 \plotone{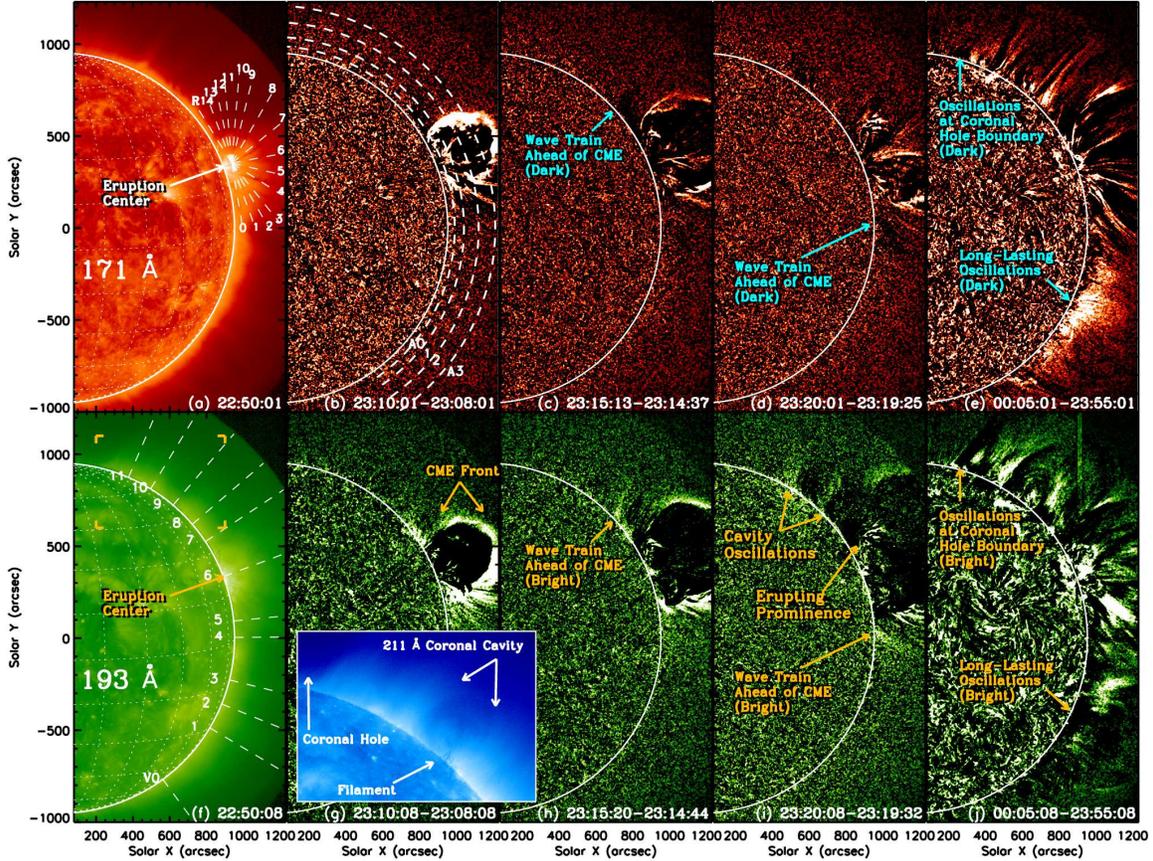}		
 \caption[]{
 AIA images 	
 at 171 (red, top) and 193~\AA\ (green, bottom),	
 with original on the left (see Movies~1A and 1B)
 and running difference on the right (see Movie~1C for 193~\AA). 
 The dashed lines indicate radial cuts R0--R14 from the eruption center (a),
 azimuthal cuts A0--A3 (b), and vertical cuts V0--V11 (f) defined in Appendix~A.
 Note EUV wave trains	
 ahead of the CME front and oscillatory brightenings and darkenings
 anti-correlated at the two wavelengths.	
 The four brackets in (f) marks the FOV of the enlarged inset at 211~\AA\ (blue)
 showing the coronal cavity and polar coronal hole.		
 } \label{mosaic_overview.eps}
 \end{figure*}
%
This event was observed by \sdoA/AIA on 2010 September 8--9 in the NOAA active region (AR) 11105
on the northwest limb. Around 22:40~UT (see Movies~1A and 1B), 
the active region core starts gradual bulging,
accompanied by the activation of a prominence that appears dark in EUV.
At about 23:05~UT, the bulging evolves into a rapid expansion
and leads to a CME expulsion. 
Simultaneously, the prominence erupts and unfolds itself, exhibiting unwinding twists 
\citep[described for other events by, e.g.,][]{LiuW.filmnt.2009ApJ...698..632L, LiuW.CaJet1.2009ApJ...707L..37L, 
LiuW.CaJet2.2011ApJ, ThompsonW.rotate-promin.2011SoPh..tmp..377T}.

A C3.3 class flare occurs at 23:00~UT, 
according to the \goes 1--8~\AA\ soft X-ray flux, peaks at 23:33~UT,
and then undergoes a slow decay for several hours (see \fig{flare_lc.eps}).
\hsi had a nighttime data gap during the flare.	
We thus use the spike of the \goes flux time derivative to define 
the {\it impulsive phase}, 23:05--23:30~UT. 		
It is during this phase that a variety of phenomena, 
including the rapid CME acceleration 
\citep{ZhangJie.CME-flare-C1.2001ApJ...559..452Z, TemmerM.synch.CME-acc.HXR.2008ApJ...673L..95T}
and the generation of the global EUV wave 
and quasi-periodic wave trains, take place.

By 23:10~UT, a dome shaped CME front,
identified as a bright rim surrounding a dimming region
\citep[e.g.,][]{ThompsonB.dimming.2000GeoRL..27.1431T},
has become evident in running difference images (see \fig{mosaic_overview.eps}). 
A few minutes later, a comparably weak intensity front appears in the {\it low} corona 
{\it ahead of} the CME flanks to the north and south.	
This weak, broad pulse, identified as a global EUV wave, contains quasi-periodic wave trains
and propagates 	
about 2--4 times faster than the lateral expansion of the CME.	
The northward EUV wave travels through a coronal cavity until 
it reaches the polar coronal hole boundary, 
instigating oscillatory brightenings.	
{\it Behind} the CME, additional quasi-periodic	
wave trains \citep{LiuW.FastWave.2011ApJ...736L..13L}
emanate from the flare  	
along a funnel of coronal loops and fade away while approaching the CME front.



Major milestones of the event are listed in \tab{table_timeline}.
We employed space--time diagrams, as detailed in Appendix~A, to analyze various features,
using four sets of narrow cuts (slits) shown in \figs{mosaic_overview.eps} and \ref{tslice_fast.eps}.
These cuts essentially form a user-defined coordinate system 
to sample images from complementing perspectives.
The {\bf eruption center} is defined at a height of $49 \arcsec$ ($1 \arcsec= 0.731 \Mm $, $R_\sun=952 \arcsec$) 
above the brightest flare kernel at N$20.6\degree$ on the limb.


\section{Global EUV Wave containing Quasi-periodic Wave Trains}		
\label{sect_decouple}			

\begin{figure}[thbp]      
 \epsscale{0.9}
 \plotone{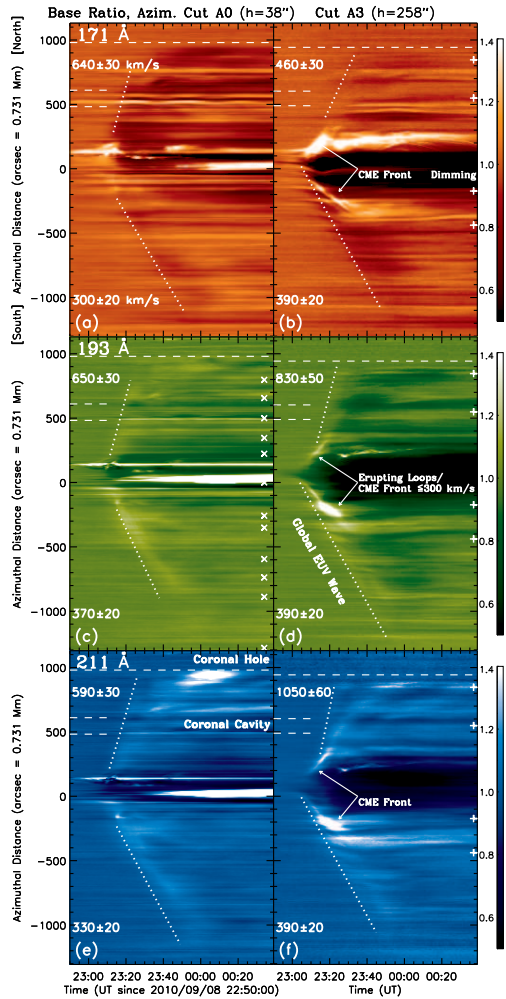}		
 \caption[]{	
 Base ratio space--time diagrams at 171 (red), 193 (green), and 211~\AA\ (blue)
 from selected azimuthal cuts A0 and A3 defined in Figure~\ref{mosaic_overview.eps}(b). 
 The distance is measured along each cut in the counter-clockwise direction 
 from the eruption center shown in Figure~\ref{mosaic_overview.eps}(a). 
 Distances shown here are normalized to that on the lowest cut A0
 to represent the azimuthal (latitudinal) position,
 a convention taken throughout this paper,
 while velocity measurements were based on the original distances.	
 %
 The dotted lines are linear fits to the global EUV wave front,
 labeled with fitted velocities (in $\kmps$). 
 The horizontal dashed lines delineate the boundaries within each cut 
 of the coronal cavity and polar coronal hole,
 best seen as dark areas in the 211~\AA\ inset of Figure~\ref{mosaic_overview.eps}.
 The cross and plus signs on the right indicate the azimuthal positions of the vertical cuts 
 defined in \fig{mosaic_overview.eps}(f) and of the temporal intensity profiles
 shown in Figure~\ref{time_prof.eps}, respectively.
 } \label{tslice_overview.eps}
 \end{figure}

\fig{tslice_overview.eps} gives an overview of the global EUV wave
seen in space--time diagrams from azimuthal cuts~A0 and A3.		 
The general properties of the wave
are examined in Appendix~B. Here we summarize the key result:
(1)~{\it Temperature variations} (Appendix~B1).
In general, the EUV wave appears as a broad stripe of darkening at 171~\AA\ 
and brightening at 193 and 211~\AA, indicating prompt heating
of plasma from 0.8 to 2.0~MK followed by restoring cooling, 
likely caused by adiabatic compression and subsequent rarefaction.
(2)~{\it Kinematics} (Appendix~B2).
Measured at 193~\AA\ on the lowest cut A0 at height $h=38 \arcsec$,
the leading front of the EUV wave travels with general deceleration
at an average velocity of $650 \pm 30 \kmps$ to the north
and $\sim$50\% slower at $370 \pm 20 \kmps$ to the south.
We use such velocities to characterize the global EUV wave,
because the 193~\AA\ wave generally signals the first arrival of the disturbance
and no systematic height dependence of the wave velocity has been found.
(3)~{\it Height dependence} (Appendix~B3). There is a clear trend of delayed arrival of the EUV wave at lower heights in the low corona 
($h \lesssim 220 \Mm$) near the eruption. This indicates a wave front		
forwardly inclined toward the coronal base.		

\subsection{EUV Wave Generation and Decoupling from CME}
\label{sect_decouple:wave-generation}

\begin{figure*}[thbp]      
 \epsscale{1.}
 \plotone{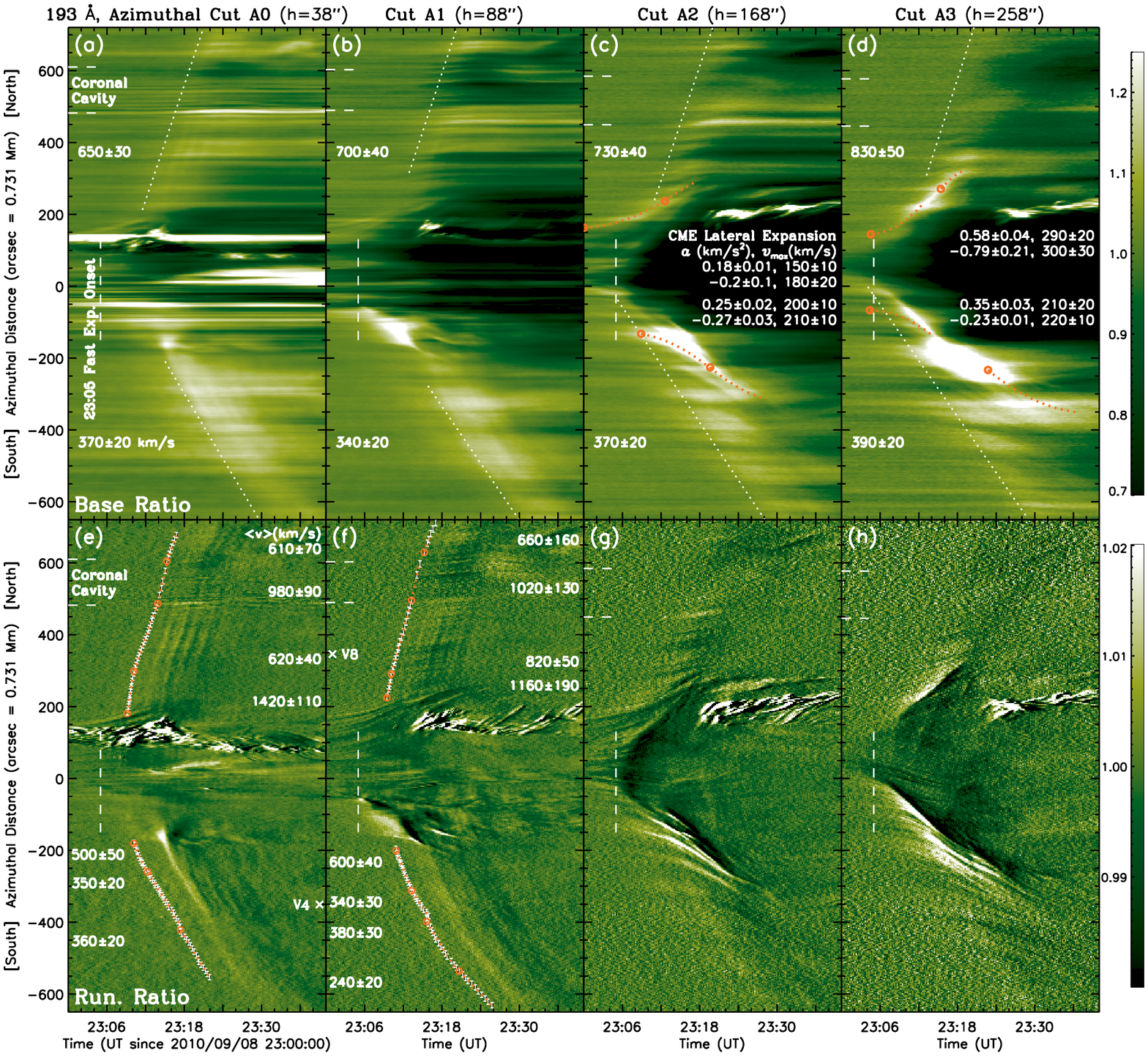}		
 \plotone{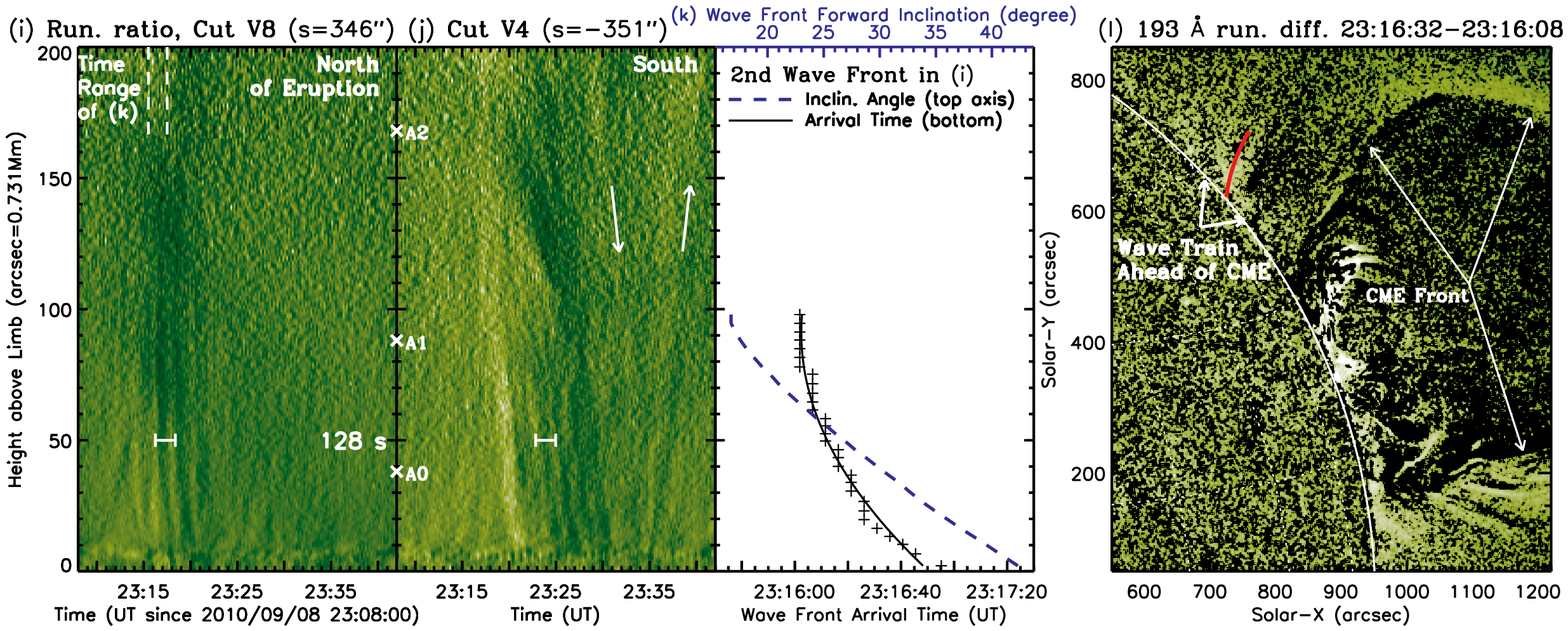}		
 \caption[]{
 Enlarged 193~\AA\ version of \fig{tslice_overview.eps} near the eruption from all azimuthal cuts
 in base ratio (top) and running ratio (middle).
 The red dotted lines in (c) and (d) are parabolic fits to the lateral CME front,	
 with fitted accelerations (in $\kmpss$) and maximum velocities (in $\kmps$) 
 listed in the order of the start time of the fit (marked by an open circle).
 Note the acceleration followed by deceleration, making the CME 	
 fall behind the global EUV wave (white dotted lines).
 The vertical dashed line marks the onset of the flare impulsive phase 
 and the rapid CME expansion at 23:05~UT.
 Note in (e) and (f) the quasi-periodic wave trains in the low corona 
 to the north and south of the eruption, in which the white data points, shifted to the left by 5~min,
 are the positions of the most pronounced wave pulses,
 fitted with piecewise linear functions (red dotted lines)		
 labeled by resulting velocities (in $\kmps$).
   {\bf Bottom}: (i) and (j) Enlarged running ratio version of \fig{tslice_vcut.eps} 
 for vertical cuts V8 and V4,	
 showing delayed arrival at lower heights of the wave trains.
   (k) Arrival time (bottom $x$-axis) at cut V8 of the second wave front in (i)
 versus height ($y$-axis). 	
 The blue dashed line (top $x$-axis) is the forward inclination angle from the local vertical
 of the reconstructed wave front, which is shown in red in the running difference
 image of (l).
 } \label{tslice_QFP.eps}
 \end{figure*}

To probe the origin of the EUV wave,	
let us examine the enlarged space--time diagrams from all azimuthal cuts in \fig{tslice_QFP.eps}.
The base ratio (top row) best shows the broad global EUV wave and post-CME dimming,
while the running ratio (middle row) highlights fast varying features,
such as the expanding CME/loops and quasi-periodic wave trains 	
(see \sect{sect_QFP:global}).

On the lowest cut A0, the global EUV wave is first detected
some $s_0=150 \arcsec = 110 \Mm$ away to the north and south of the eruption center,
but at different times. The northward wave appears at 23:11:20~UT,
6~min after the common onset of the fast CME expansion and flare impulsive phase,
as described in Appendix~C. By that time, the CME front to the north near the vertical
direction has reached $>$$210 \kmps$. The southward wave is generated 3~min later
at the outermost edge of some low-lying loops rapidly pushing downward (\figs{tslice_radial.eps}(m) and (n)). 
Within the distance $s_0$, no systematic emission change can
be related to either the EUV wave or CME expansion.

On higher cuts A1--A3 (\fig{tslice_QFP.eps}), while the EUV wave signal decreases,
loop expansion due to the upwardly growing CME becomes increasingly evident.
On cuts A2 and A3, the EUV wave can be traced further back in distance 
to near the eruption center and back in time 
as the successive onsets of the CME's rapid lateral expansion.
This suggests that the EUV wave
could be the continuation into the quiet Sun region
of the same disturbance signaling the fast expansion
\citep[cf., differential expansions in other events;][]{Schrijver.Elmore.filament-acc.2008ApJ...674..586S,
Schrijver.Aulanier.2011Feb15.X2.AIAwv.2011ApJ...738..167S}.
%
%

This rapid CME expansion may play a key role in driving the EUV wave,
as described in Appendix~C. It involves a sizable volume 
of roughly the active region of $r_{\rm AR}= 110 \Mm$ in radius and 
originates at its center of a height $h_0=110 \Mm$ above the photosphere.
Below this height, the CME expansion has a significant downward component,
while the lateral expansion is negligibly slow and spatially confined
(see \fig{tslice_radial.eps} and Movie~1C). 
Above this height, as shown in \figs{tslice_QFP.eps}(c) and (d), the lateral CME expansion experiences
an acceleration phase with $a=(0.18 \pm 0.01)$--$(0.58 \pm 0.04) \kmpss$ 
and maximum velocities $(150 \pm 10)$--$(300 \pm 30) \kmps$, 
followed by 	
a comparable deceleration. 
This makes the CME flanks quickly fall behind the global EUV wave 	
that is 2--4 times faster, consistent with recently reported CME-wave decouplings
\citep{Patsourakos.EUVI-fast-mode-wave.2009ApJ...700L.182P, Kienreich.2009ApJ...703L.118K,
ChengX.ZhangJie.EUV-wave-CME-decouple.2012ApJ...745L...5C}.
%

The 110~Mm distance of the initial EUV wave from the eruption center
is comparable to those of other EUV waves
\citep[e.g.,][]{Veronig.STEREO-EUVI-wave.2008ApJ...681L.113V, LiuW.AIA-1st-EITwave.2010ApJ...723L..53L}
and Moreton waves or type~II bursts \citep{WarmuthA.multiw-EIT-wave.2004A&A...418.1101W,
MuhrN.2003Oct28.Moreton.wave.2010ApJ...708.1639M},
suggesting a spatially extended, rather than a point-like source region.
Whether the EUV wave represents an MHD wave or CME caused restructuring,
such a considerable distance, 	
in general, is required for the velocity and volume of the expanding CME (as the disturbance driver)
to grow, in order to build up sufficient compression to 	
the ambient corona to produced detectable emission changes.		
In addition, the associated 6--9~min delay 
\citep[cf., 1--2~min,][]{VrsnakB.shock-delay_1-2min.1995SoPh..158..331V}
seems reasonable for a piston-driven, nonlinear wave to steepen into a shock
\citep{MannG.shock-formation.1995JPlPh..53..109M, VrsnakB.Lulic.coronal-shock-1.2000SoPh..196..157V},
because it is on the same order of the predicted shock formation time
\citep[][see, e.g., their Figure~6(a)]{Zic.piston-driver-shock.2008SoPh..253..237Z}
by applying relevant control parameters from our measurements.



\subsection{Quasi-periodic Wave Trains Ahead of CME}
\label{sect_QFP:global}

The quasi-periodic wave trains ahead of the CME appear as successive arc-shaped
pulses in the low corona	
in running difference images (see \figs{mosaic_overview.eps}(h) and (i), Movie~1C). 
They are best seen in running ratio space--time diagrams (\fig{tslice_QFP.eps}, middle)
as fine-structure, narrow stripes of alternating brightening and darkening,
within the broader pulse of the global EUV wave seen in base ratio (top).
The northward wave train is particularly evident, 
including coherent pulses 	 
traveling some $500 \arcsec$ ($\gtrsim$$R_\sun/2$) away.
The first pulse in each direction is located at the leading front of the global wave,
appearing at $s_0= 150 \arcsec$ from 	
the eruption center 		
ahead of the CME flank,	
followed by subsequent pulses originating from about the same location.
Each pulse has the same wavelength dependence
as the overall global wave noted above (see Appendix~B1), indicating plasma heating.


The sharpness of these narrow pulses allowed us to trace detailed kinematics 
better than the diffuse front of the global EUV wave.
We selected the most pronounced pulse from cuts A0 and A1 in each direction, 
i.e., the second in the north and the first in the south. 
We then fitted their space--time positions with a piecewise linear function,
as shown in red in \figs{tslice_QFP.eps}(e) and (f), shifted by $-5$~min to allow
the original data visible.  
On cut A0, the velocities within the first 1--2~minutes are
$1420 \pm 110$ and $500 \pm 50 \kmps$ in the north and south, respectively,
and then drop by $\sim$50\% to the average velocities 	
of the diffuse wave front shown in panel~(a).	
For the northward pulse, 	
there is an additional velocity change		
upon its arrival at the coronal cavity, in which it travels at
$980 \pm 90 \kmps$, about 50\% faster than the $620 \pm 40 \kmps$ velocity outside,
followed by a drop back to this level upon its exit from the cavity's far side. 
Such non-monotonic velocity variations resemble those in other EUV waves 
under different circumstances \citep{Zhukov.EIT-wave-acc-decel.2009SoPh..259...73Z,
LiTing.2010Jun07.AIAwave.reflect.2012ApJ...746...13L}.
A similar behavior occurs on cut A1, but we found no systematic
height dependence of the wave velocity.


The height distribution of the quasi-periodic wave trains can be seen in running ratio
space--time diagrams from vertical cuts V8 (northward) and V4 (southward), 
as shown in \figs{tslice_QFP.eps}(i) and (j).
These waves are prominent in the low corona $h< 150 \arcsec \approx 110 \Mm$
and their signals rapidly fall with height. At least five pulses (stripes), uniformly spaced in time by about
2 min (128~s) can be identified in each direction.
These stripes have negative slopes, indicating delayed arrival 
at progressively lower heights and wave fronts forwardly inclined toward the solar surface,
same as the overall global EUV wave noted above (see Appendix~B3). 
To the south, the shallower slope is consistent with the lower propagation velocity
and thus longer delays.	The sudden change of slope from negative to positive
at 23:35~UT is a possible indication of reflection of the wave at the coronal base.

The arrival time of the second pulse from cut V8, for example, is plotted against height ($y$-axis)
in \fig{tslice_QFP.eps}(k) and fitted with a parabolic function (solid line), indicating a delay of $53 \s$ over 
a height range of $98 \arcsec$. As detailed in Appendix~B3, we backtracked in time this delayed arrival
to reconstruct the wave front seen in the running difference Movie~1C
that is, however, difficult to identify in stills.
The result is shown in red in panel~(l).
Its forward inclination angle from the local vertical is shown in blue in panel~(k),
which increases with decreasing height, ranging from $17\degree$ at $h=98 \arcsec$ 
to $42 \degree$ at the coronal base.

\section{Sequential Oscillations on Wave Path}
\label{sect_oscil}

 \begin{figure}[thbp]      
 \epsscale{0.95}
 \plotone{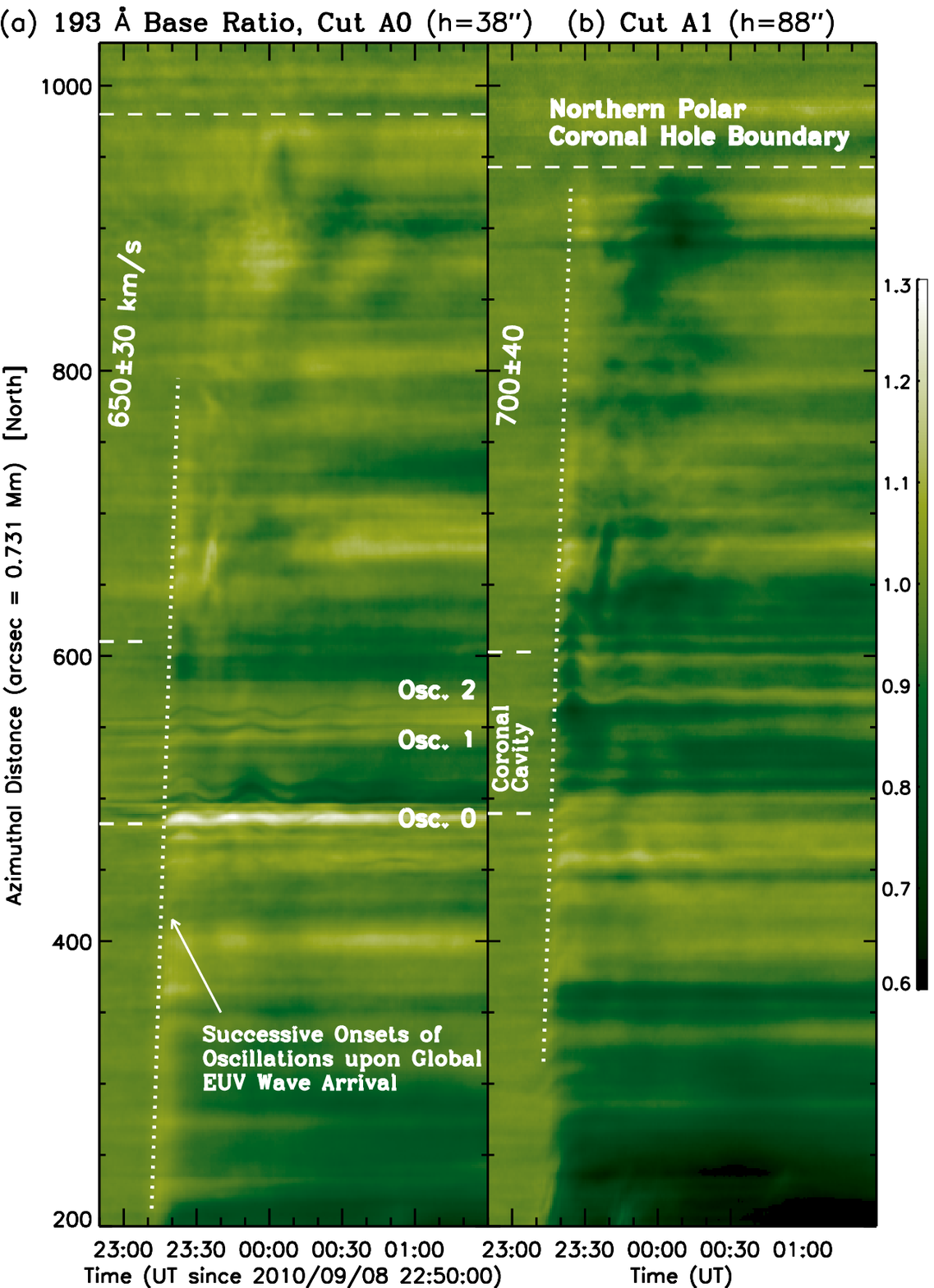}		
 \epsscale{0.9}
 \plotone{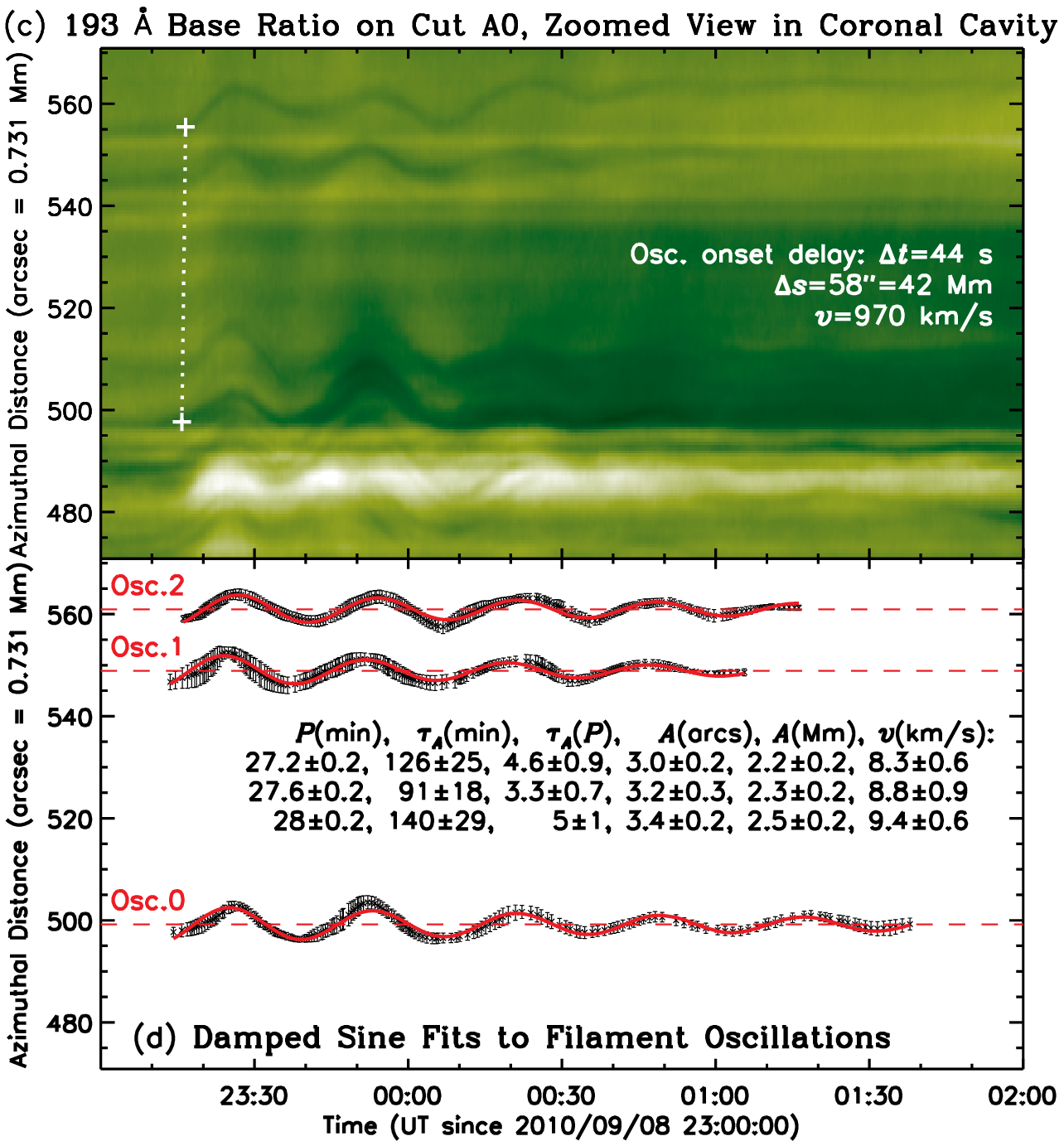}		
 \caption[]{
 (a) and (b) 193~\AA\ base ratio space--time diagrams along azimuthal cuts A0 and A1
 to the north of the eruption, showing an uninterrupted sequence of transverse oscillations set off 
 by the arrival of the global EUV wave at increasing distances.	
 (c) Zoomed view of (a) showing filament oscillations within the coronal cavity,
 fitted with damped sine functions in (d). Note in (c) the delayed onsets of oscillations at 
 greater distances in agreement with the EUV wave travel time at the measured velocity.
 } \label{tslice_track_oscil.eps}
 \end{figure}
 \begin{figure*}[thbp]      
 \epsscale{0.9}
 \plotone{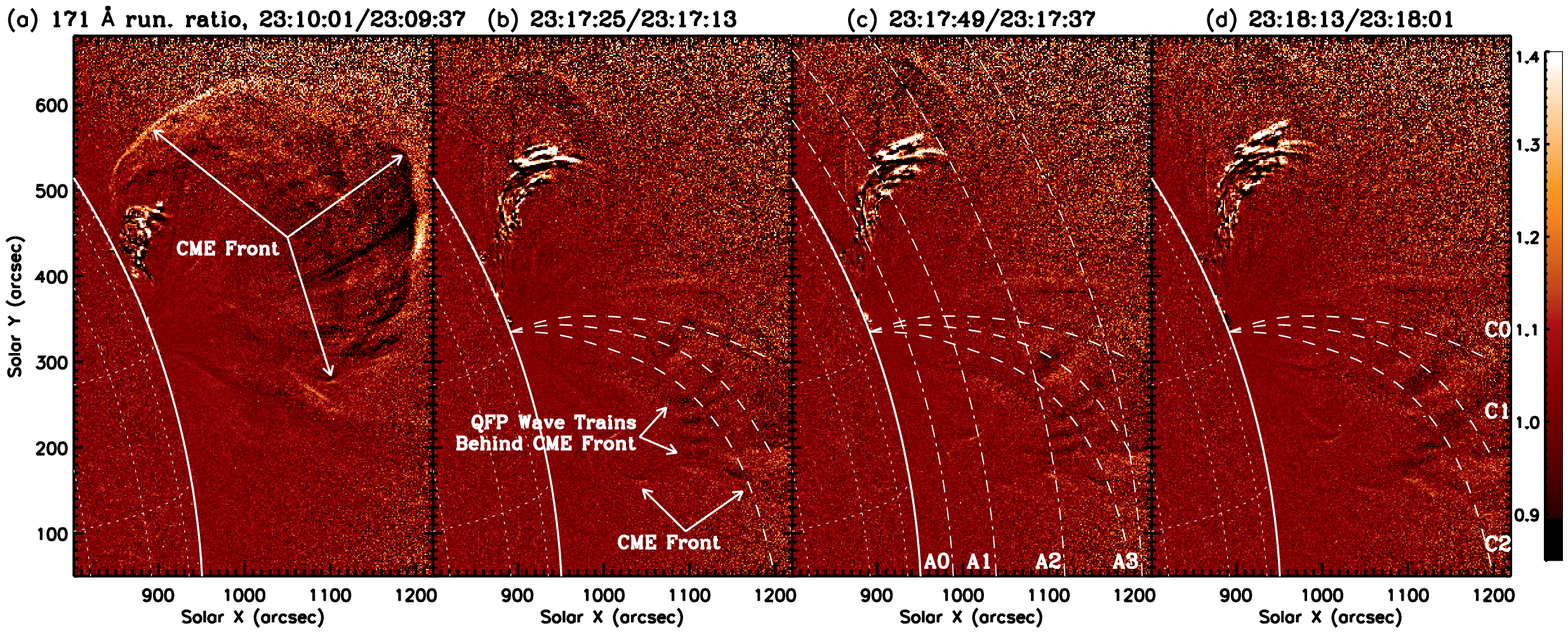}		
 \epsscale{0.9}
 \plotone{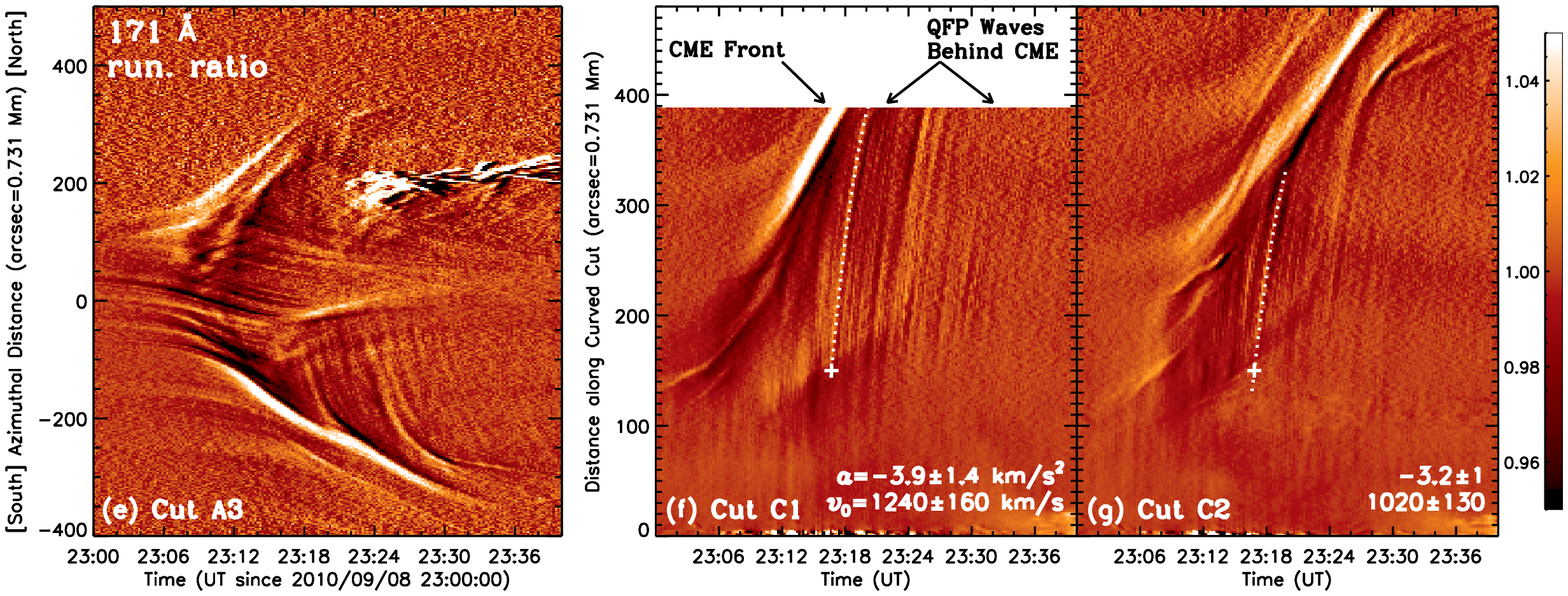}		
 \caption[]{
 QFP wave trains seen at 171~\AA\ in the funnel rooted at the flare kernel behind the CME front.
   {Top}: Running ratio images (see Movies~2A for original
 and 2B for running difference).
 The dashed (dot-dashed) lines mark the curved cuts C0--C2 (azimuthal cuts A0--A3).
   {Bottom}: Running ratio space--time diagrams from selected azimuthal cut A3 
 and curved cuts C1 and C2.
 Overlaid dotted lines are parabolic fits to a single wave front seen in the
 respective cuts, labeled with fitted deceleration $a$ and initial velocity $v_0$ at a
 reference distance of $s_0= 150 \arcsec$ (plus sign).	
 } \label{tslice_fast.eps}
 \end{figure*}
%
As mentioned earlier, upon the arrival of the global EUV wave 
at increasing distances, local coronal structures are sequentially deflected,
setting off transverse oscillations, which are best seen as sinusoidal stripes in enlarged
base-ratio space--time diagrams (\fig{tslice_track_oscil.eps}).	
In running-ratio diagrams (\figs{tslice_QFP.eps}, middle), they appear as numerous
shallow-sloped, narrow stripes superimposed on the steep, broad pulse of the global EUV wave, 
as noted in an earlier event \citep[][see their Figure~5(d)]{LiuW.AIA-1st-EITwave.2010ApJ...723L..53L}.
We found oscillations lasting up to 6~hr in a variety of structures, 
including the northern polar coronal hole boundary,
with velocity amplitudes of 8--$27 \kmps$ 
and periods ranging from 12~min in short loops
to 80~min in large-scale loops.
We expect to use such high resolution measurements 		
as seismological tools in the future to infer 
physical conditions and damping mechanisms
\citep[e.g.,][]{Ofman.Aschwanden.damping.scaling.2002ApJ...576L.153O,
Ofman.oscil.model-review.2009SSRv..149..153O,
WhiteRS.Verwichte.AIA.loop.oscil.2012A&A...537A..49W, 
WangTJ.AIA.transverse.oscil.2012arXiv1204.1376W}.
Here we focus on the best manifestation of such oscillations
in the coronal cavity 		
to shed light on the nature of the EUV wave.

\subsection{Oscillations of the Coronal Cavity}
\label{sect_oscil:cavity}


We selected three dark filament threads (labeled Osc.~0, 1, and 2)
embedded in the coronal cavity as tracers of its oscillations seen
in \fig{tslice_track_oscil.eps}.	
We obtained the space--time position $s(t)$ at the center of each thread,
using its width as the spatial uncertainty. 
$s(t)$ was smoothed to approximate the slowly drifting
equilibrium position $s_{\rm eq}(t)$, which was then removed to 
obtain the net oscillatory displacement $\Delta s(t) =s(t) - s_{\rm eq}(t) $,
as shown in \fig{tslice_track_oscil.eps}(d).
We fitted $\Delta s(t)$ with a damped sine function,
$\Delta s(t) =  A \exp [- (t-t_0) / \tau_A ] \sin [ 2 \pi (t- t_0) / P ]$,
shown as the red solid line, where $A$ is the initial amplitude at $t=t_0$, $P$ the period,
and $\tau_A$ the e-folding damping time.
	%
	%
	%
These filament threads, although located at different positions,	
have fitted parameters in a narrow range,
indicating a coherence consistent with the bodily oscillations of the entire
cavity.		
The average parameters are
$\langle \tau_A \rangle= (119 \pm 14) \min$,		
$\langle P \rangle=(27.6 \pm 0.1) \min$,
and $\langle A \rangle=3 \farcs 2 \pm 0 \farcs 1 = (2.3 \pm 0.1) \Mm$,
giving an initial velocity amplitude
$\langle v \rangle = \langle 2 \pi A / P \rangle = (8.8 \pm 0.4) \kmps $.


The onsets of oscillations within the cavity further away from the eruption
are delayed in time.		
For example, Osc.~2 at $\Delta s = 58 \arcsec = 42 \Mm$ from
Osc.~0 is delayed by $\Delta t = 44 \s$.
This translates to a velocity of $\Delta s / \Delta t = 970 \kmps$,
in agreement with the elevated velocity of the wave trains within
the cavity found earlier (see \fig{tslice_QFP.eps}(e)).

As detailed in Appendix~D, we estimated the energies of the cavity oscillations and global EUV wave 
to be $10^{27}$ and $10^{28} \ergs$, respectively, significantly lower than 
the $10^{31} \ergs$ kinetic energy of the CME. 
The transverse oscillations of the cavity are likely fast kink modes.
Its magnetic field was thus estimated via coronal seismology 
to be on the order of $\sim$6~G with a pitch angle of $\sim$$70 \degree$,
suggestive of a highly twisted flux rope.

\section{Comparing Two Types of Quasi-periodic Wave Trains}
\label{sect_QFP}


We now examine the quasi-periodic wave trains within the funnel {\it behind} the CME
and flare pulsations, to help understand the wave trains in the global EUV wave {\it ahead of} the CME
presented in \sect{sect_QFP:global}.
%
 \begin{figure*}[thbp]      
 \epsscale{1.}
 \plotone{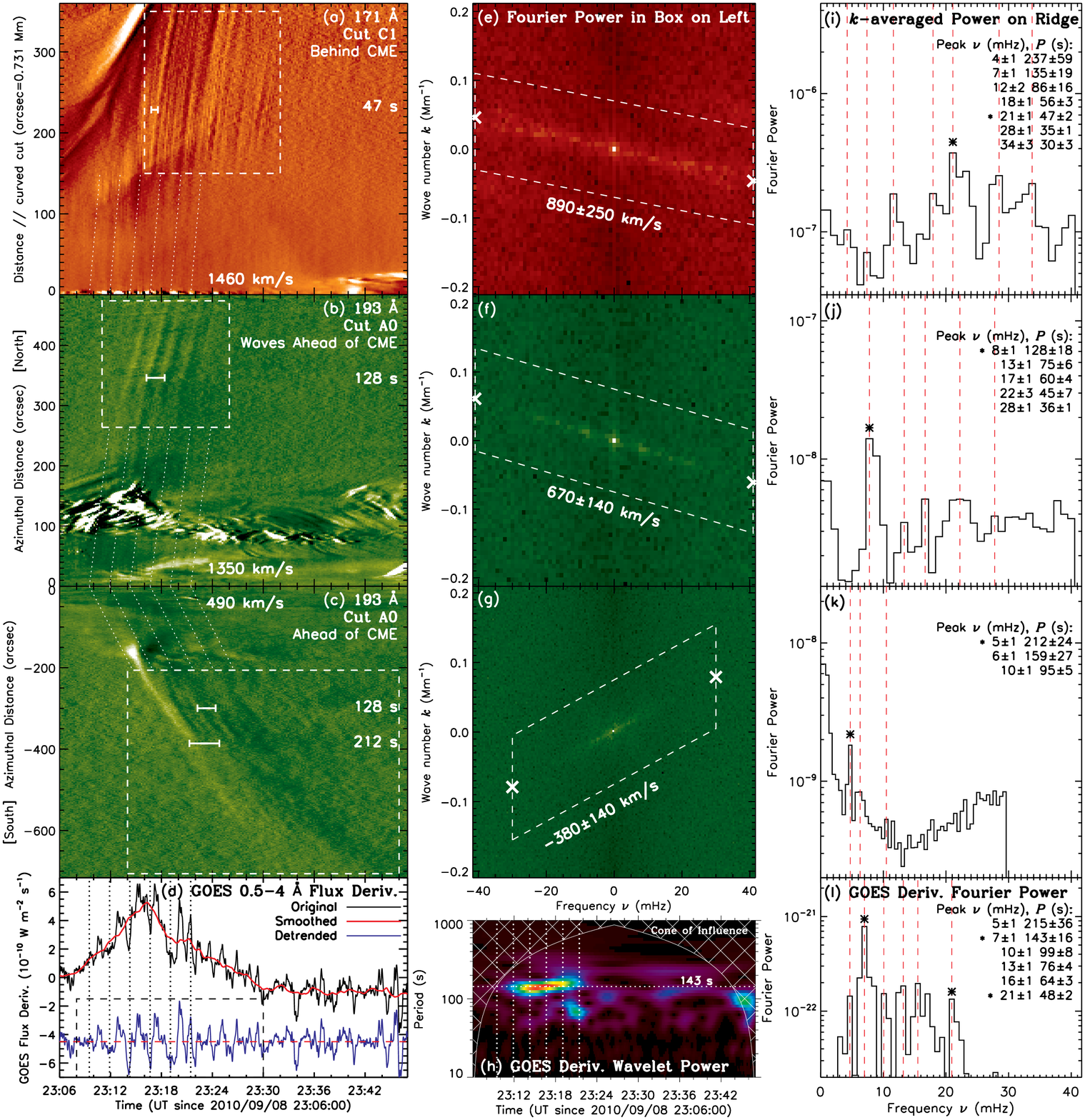}		
 \caption[]{
  Fourier analysis of quasi-periodic waves and flare pulsations.
  {\bf Top three rows:} 
  {\it Left}: Running ratio space--time diagrams from curved cut C1 at 171 \AA\ (a) and azimuthal cut A0 at 193 \AA\
 to the north (b) and south (c) of the eruption, showing waves
 behind and ahead of the CME, respectively.
  {\it Middle}: Fourier power ($k$--$\omega$ diagram) on logarithmic scales 
 of the data cube 	
 within the boxed region on the left. The cross signs mark the end points of a power-weighted linear fit 
 to the bright ridge.
  {\it Right}: Fourier power averaged over wave number $k$ within the parallelogram in the middle. 
 The frequencies and periods of peaks identified by red vertical lines are listed on the right. 
 Selected prominent peaks are marked by asterisks and their periods are indicated by 
 the horizontal bars on the left.
  {\bf Bottom row:} 
 (d) Time derivative of the \goes 0.5--4~\AA\ flux. 
 Subtracting the gradual, smoothed data (red solid line) gives the detrended blue curve,
 whose wavelet power and Fourier power 
 (within the time range defined by the dashed box) are shown in (h) and (l), respectively.
 The vertical dotted lines mark five pulsations at the dominant 143~s period.
 } \label{fast-fft.eps}
 \end{figure*}
%

\subsection{QFP Wave Trains in Funnel Behind CME}
\label{sect_QFP:inCME}


	%
\fig{tslice_fast.eps} (top panels, see Movies~2A and 2B)
shows the quasi-periodic wave trains behind the CME 	
as successive arc-shaped intensity variations on the 1--5\% level,
traveling along a funnel of coronal loops rooted at the flare kernel
and curved toward the south. 
No significant signals can be detected in channels other than 171~\AA,
suggesting that the responsible plasma is likely near 
this channel's peak response temperature of $0.8 \MK$.
In space--time diagrams (bottom panels),	
these waves are steep, recurrent stripes behind the bright CME front. 
Parabolic fits (dotted lines) to a prominent wave pulse in different directions
indicate large decelerations of 3--4~$\kmpss$, 
with initial velocities at $s_0=150 \arcsec$ from the flare kernel
of $(1020 \pm 130)$--$(1240 \pm 160) \kmps$,		
which are comparable to those of the northward quasi-periodic waves ahead of the CME, but twice
those of the southward waves (see \fig{tslice_QFP.eps}(e)).

We identify the wave trains in the funnel as quasi-periodic fast propagating (QFP) waves
recently discovered by AIA \citep{LiuW.AIA-1st-EITwave.2010ApJ...723L..53L, 
LiuW.FastWave.2011ApJ...736L..13L, LiuYu.2012ApJ.QPF.waves}
and confirmed as fast-mode magnetosonic waves in 3D MHD simulations 
\citep{Ofman.Liu.fast-wave.2011ApJ...740L..33O}.
The new characteristics found here are  	
the rapid deceleration and disappearance while approaching the CME front,
which will be discussed in \sect{subsect_conclude:proposal}.


\subsection{Fourier Analysis of Wave Trains Behind/Ahead of CME	
and Flare Pulsations}
\label{sect_QFP:FFT}

We now apply Fourier analysis to the quasi-periodic waves both ahead of 
and behind the CME front, and compare them with flare pulsations.
From each of the space--time diagrams shown in \figs{fast-fft.eps} (left), 
we extracted a data cube from the dashed box in which the wave signals are prominent.
We then obtained its Fourier power ($k$--$\omega$ diagram) in the domain of 
temporal frequency $\nu$ and spatial wavenumber $k$, as shown in the middle.
Each $k$--$\omega$ diagram displays a bright ridge that represents the dispersion relation
of the wave averaged over the extracted duration. 
Within a selected region (dashed line) around the ridge,
the $(\nu, k)$ positions of bright pixels above a threshold (10\% of the maximum),
weighted by the power, were fitted with a linear function.
These fits (passing through the origin),
as marked by the two cross signs on the sides,	
indicate average phase velocities of $v_{\rm ph}= 890 \pm 250, \, 670 \pm 140, \, -380 \pm 140 \kmps$,
which agree with the corresponding velocities measured in the space--time diagrams
(see \figs{tslice_QFP.eps} and \ref{tslice_fast.eps}).
Here the $+$/$-$ signs denote the propagation direction and the
uncertainties imply scatter due to temporal variations.	


By averaging Fourier power over wave number $k$
within the selected regions, we obtained the frequency distributions
of these waves (\figs{fast-fft.eps}, right).
Prominent peaks were fitted with Gaussians to find the corresponding frequencies (periods)
and their $1 \sigma$ uncertainties.
As a general trend, the QFP waves behind the CME front are preferentially
distributed at high frequencies, with the highest peak 
(marked by an asterisk) located at $\nu = 21 \pm 1$~mHz ($P=47 \pm 2 \s$),
while the waves ahead of the CME are concentrated at low frequencies, 
with the highest peaks at $128 \pm 18 $ and $212 \pm 24 \s$ for the north- and southward waves, respectively.
The second highest peak of the southward waves is at $159 \pm 27 \s$,
similar to the highest peak of the northward waves.	
We thus refer to both with a {\bf 2~min periodicity}. 
Such periodicities can also be visually identified in the space--time 
domain on the left.

In an earlier event \citep{LiuW.FastWave.2011ApJ...736L..13L},
a 3~min period was found in both the QFP waves in a funnel
and X-ray and UV flare pulsations
\citep{OfmanSui.HXR-oscil.2006ApJ...644L.149O, Nakariakov.Melnikov.QPP.2009SSRv..149..119N}.
We searched for such a common periodicity in this event.
Because of no hard X-ray coverage and limb occultation of
UV ribbons here, we used the time derivative of the \goes 0.5--4~\AA\ soft X-ray flux
as a proxy of hard X-rays,	
assuming the \citet{NeupertW1968ApJ...153L..59N} effect at work \citep{ 		
VeronigA2005ApJ...621..482V, LiuW2006ApJ...649.1124L,	
NingZJ.CaoWD.Neupert-type.2010ApJ...717.1232N}.		
As shown in \fig{fast-fft.eps}(d), there are clear pulsations 
superimposed on a slower variation 
shown in red, which was then removed to give a detrended light curve shown in blue.
%
We extracted the detrended data within the interval defined by the dashed box,
covering most of the impulsive phase, 	
and obtained its Fourier power (panel~(l)). 
The power is primarily distributed within 3--23~mHz, 
with concentration at low frequencies or long periods (2--3.5~min). 
This trend is similar to that of the EUV waves ahead of the CME (panels~(j) and (k)),
but differs from that of the QFP waves behind the CME with stronger power
at high frequencies (panel~(i)). 
In particular, the prominent flare period of $143 \pm 16 \s$ 	
agrees, within uncertainties, with the 2~min periodicity of the waves ahead of the CME.
The wavelet power (panel~(h)) of the flare light curve	
shows a peak of such pulsations,	
five of which, marked by the vertical dotted lines, can be identified in panel~(d).
On the other hand, despite their different general trends, the flare pulsations
and the waves behind the CME share some common frequencies within their
broad distributions, including a prominent peak at $21 \pm 1$~mHz or $47 \pm 2 \s$.

We now return to the space-time domain to further explore the connections
between the flare pulsations and the two groups of quasi-periodic EUV waves.
In each space-time diagram of \fig{fast-fft.eps} (left),
we located the onset times of the first identified flare pulsation
at the origin ($s=0$, the flare kernel)		
and of the quasi-periodic waves at the distance of the near edge 
of the box. We then connected the two positions
with a straight dotted line, followed by parallel lines 
corresponding to the rest of the five pulsations at the $143 \s$ period.
Interestingly, these slanted lines translate to velocities of $1460$, $1350$, and $490 \kmps$
that agree with the initial velocities of the respective wave trains
noted earlier (see \figs{tslice_QFP.eps} and \ref{tslice_fast.eps}).
In addition, for the northward wave train ahead of the CME, the slanted lines closely trace
the first two pulses and match the onsets of the rest of the five
pulses at the edge of the box. 
For the southward waves, such a close match is less evident,	
but the duration of these flare pulsations extrapolated to the box edge		
overlaps with that of the quasi-periodic waves there.
Implications of these coincidences will be discussed in \sect{subsect_conclude:discuss}.
%



\section{Concluding Remarks}	
\label{sect_conclude}

\subsection{Summary}
\label{subsect_conclude:Summary}

We have presented detailed analysis of a global EUV wave containing quasi-periodic wave trains
ahead of the CME front,	
focusing on three interrelated aspects:
the generation of the EUV wave and its decoupling from the CME,
sequential structural oscillations (particularly of the coronal cavity) triggered by
the wave, and a comparison of these wave trains with those behind the CME front and 
with flare pulsations. We summarize the general properties of various phenomena 
in \tab{table_summary} and recapitulate below our findings,	
the first two being reported for the first time:		
%
\begin{table}[bthp]	
\scriptsize	
\caption{General Properties of Phenomena Involved in the Event.}
\tabcolsep 0.03in	
\begin{tabular}{lrrrrrrrrrrr}
\tableline \tableline

Phenomena            & \multicolumn{3}{c}{$v$ ($\kmps$)} & $Period$ & $E$($\ergs$) &  \\

\tableline \tableline

CME             & \multicolumn{2}{l}{lateral 150--$300$} & vertical 990 & & $10^{31}$   \\
Flare                                              & & & & 2 min &  \\

\tableline
\multicolumn{5}{l}{Wave Trains (WTs) ahead of CME (Global EUV Wave):}  & $10^{28}$         \\
(1) Northward    & Init.~1420 & Ave.~650  & In Cav.~980  & 2 min   \\
(2) Southward    &       500  &  370      &                 & 2 min  \\

WTs behind CME  & 1020--1240  &  &   &  47 s  \\

\tableline
Cavity Osc.  & 9          & &   & $28 \min$ &  $10^{27}$  \\

\tableline  \end{tabular}


\label{table_summary}
\end{table}
%




\begin{enumerate}	

{\bf \item	

Quasi-periodic wave trains}%
 \footnote{Such wave trains were somewhat discernible in an earlier \stereo event at a lower cadence of 75~s,
  but were not analyzed \citep[][see their Figure~11]{Patsourakos.early-CME_EIT-wave-train.2010A&A...522A.100P}.
  }
within the broad global EUV wave pulse travel ahead of the lateral CME front
coherently to $\gtrsim$$R_\sun/2$ along the solar surface.
Their initial velocity 			
of $1420 \pm 110 \kmps$ ($500 \pm 50 \kmps$) to the north (south) of the eruption
decelerates to their average velocity $\sim$50\% slower.		
Their dominant 2~min period agrees with that of the X-ray flare pulsations.
These wave trains are spatially confined to the low corona $h \lesssim 110 \Mm$,
similar to the heights of global EUV waves seen by \stereo 
\citep{Patsourakos.EUVI-wave.2009SoPh..259...49P, Kienreich.2009ApJ...703L.118K}.
Their arrival is delayed at lower heights by up to three minutes,
indicating wave fronts being forwardly inclined toward the coronal base
(see \figs{tslice_QFP.eps} and \ref{fast-fft.eps}, \sect{sect_QFP:global}),
while we found no systematic height dependence of the wave velocity (\fig{multi_vfit.eps}).


{\bf \item	

Sequential transverse oscillations} and deflections of local structures,
upon the arrival of the global EUV wave at increasing distances,
form an {\bf uninterrupted chain reaction}.
The best manifestation is the (likely fast kink mode) oscillations 
of the coronal cavity and its embedded filament,
with a 28~min period and $9 \kmps$ velocity amplitude.
The delayed oscillation onsets across the cavity 
agree with the travel time of the EUV wave at
an elevated velocity of $\sim$$980 \kmps$ within it.
The estimated energies of the cavity oscillations and global EUV wave are
$10^{27}$ and $10^{28} \ergs$, respectively, $10^3$--$10^4$	times	
lower than the $10^{31} \ergs$ kinetic energy of the CME (\fig{tslice_track_oscil.eps}, \sect{sect_oscil}).

\item 	

At low altitudes $h< 110 \Mm$, the global EUV wave is first detected
at $s_0=110 \Mm$ to the north and south of the eruption center,
6--9~min after the coincidental onset of the rapid CME expansion and flare impulsive phase.
At higher altitudes, the EUV wave can be traced
further back in distance to near the eruption center and back in time 
as the successive onsets of rapid loop expansions involved in the CME
(\fig{tslice_QFP.eps}, \sect{sect_decouple:wave-generation}).


\item	

The fast expansion of the CME, with a significant downward component, 	
involves a sizable volume of about the active region
of $110 \Mm$ in radius with its center at an elevated height of $110 \Mm$
(\fig{tslice_radial.eps}, Appendix~C).
Negligibly slow near the coronal base, the lateral expansion
grows with height, reaching 150--$300 \kmps$ and then decelerating,
falling behind the global EUV wave of 2--4 times faster (\fig{tslice_QFP.eps}). 

\item	

The broad global EUV wave and its narrow quasi-periodic pulses 	
appear as darkening at 171~\AA\ and brightening at 193 and 211~\AA, 
with the maximum intensity variations being generally delayed from 193 to 211 and 171~\AA,
suggestive of heating from typically 0.8 to 2.0~MK followed by restoring cooling
(\figs{tslice_overview.eps} and \ref{time_prof.eps}, Appendix~B1).

\item	

There are additional QFP wave trains
\citep[fast magnetosonic;][]{LiuW.FastWave.2011ApJ...736L..13L, Ofman.Liu.fast-wave.2011ApJ...740L..33O} 
of a shorter, dominant period of $47 \pm 2 \s$, traveling at 1000--$1200 \kmps$ along 
the coronal funnel rooted at the flare kernel {\it behind} the CME front,
near which they abruptly disappear	
(\fig{tslice_fast.eps}, \sect{sect_QFP:inCME}). 	

\end{enumerate}	


\subsection{Proposed Physical Picture}
\label{subsect_conclude:proposal}


We propose the following physical picture, as shown earlier in \fig{cartoon.eps},
to tie together these observations.
After a gradual rise and bulging of the
active region of radius $r_{\rm AR}=150 \arcsec = 110 \Mm$, 
a rapid expansion, likely driven by a Lorentz force,
initiates near its center located at a height $r_{\rm AR}$ above the photosphere.
Within minutes, this expansion	
soon involves the entire active region
of a significant volume and becomes sufficiently fast ($\gtrsim$$200 \kmps$),
leading to a CME expulsion \citep{Patsourakos.CME-genesis.AIA.2010ApJ...724L.188P}.
An EUV wave is then generated in the low corona at the outer edge of
the expanding CME, $s_0=110 \Mm$ from the eruption center.
This distance and the 6--9~min delay allow the CME expansion to grow and build up
sufficient compression to the ambient corona or allow a nonlinear wave to steepen
into a shock.

Initially centered at a significant height of $110 \Mm$,	
the expanding CME volume naturally produces the sphere-like shape of the early EUV wave front.
The elevated lateral expansion and underlying chromosphere of 
extremely high density, essentially serving as a hard floor, compose an angled formation.
Compared with the upward expansion into the open corona, this formation
produces a compression to the surrounding low corona at an enhanced efficiency.
This is similar to a piston in a semi-enclosed volume
and may play an important role in driving the low-corona portion 
of the EUV wave that is forwardly inclined to the solar surface. 
The {\it downward} component of the compression%
 \footnote{As a possibility yet to be validated,
 such a downward compression could be 	
 physically similar to recently proposed Lorentz transients
 \citep{FisherG.Lorentz.force.sunquake.2012SoPh..277...59F}
 in flares that generate photospheric magnetic field changes
 \citep{WangHM.LiuC.Back.Reaction.2010ApJ...716L.195W, SunXD.HMI.2011Feb15.X1.5.implosion.2012ApJ...748...77S}
 and even sunquakes \citep{ZharkovaV.sunquake.2011-02-15.flux-rope.2011ApJ...741L..35Z, 
 HudsonH.flare.momentum.2012SoPh..277...77H}.
 }
may cause on-disk redshifts in EUV waves 	
 \citep{Harra.Sterling.EIS.2010Feb16.EITwv.2011ApJ...737L...4H,
Veronig.EIS.2010Feb16.EITwv.2011ApJ...743L..10V} and 
magnetic transients in Moreton waves
\citep{KosovichevZharkova.Moreton-wave-MDI.1999SoPh..190..459K}.
Such compression can explain the preferential height distribution of the EUV wave 
in the {\it low} corona, and can produce density enhancements and adiabatic heating
\citep{Schrijver.Aulanier.2011Feb15.X2.AIAwv.2011ApJ...738..167S, 
DownsC.MHD.2010-06-13-AIA-wave.2012ApJ}, 
which are likely responsible for the observed emission variations,
perhaps together with other heating mechanisms.
This adds to the importance of the lateral (versus vertical) CME expansion
in EUV wave generation \citep{PomoellJ.MHD-EIT-wave-lateral-expansion.2008SoPh..253..249P,
TemmerM.analytic-model-Moreton-wave.2009ApJ...702.1343T}.


As the CME expands and its center rises, 	
the ambient corona parts to allow the CME expulsion
into the heliosphere. Its lateral expansion at increasing heights 
then slows down and halts, partly because of the resistance of the ambience
by a Lorentz restoring force. This produces a dimming region of a limited spatial extent.
However, the EUV wave	
proceeds at the local fast-mode velocity further to the rest of the corona,
leading to the spatial decoupling from its initial driver, the CME.
The periodicity (2~min) and coherence over great distances ($\gtrsim$$R_\sun/2$)
of the wave trains within the broad global EUV wave pulse	
clearly indicate their true wave nature.	
Their velocities of 300--$1400 \kmps$ 		
are also within the expected range of coronal fast-mode speeds.		
If the simulated ``coronal Moreton waves" 
\citep{ChenFP.EIT-wave-MHD.2002ApJ...572L..99C, ChenFP.EIT-wave.2005ApJ...622.1202C} 
repeatedly launched and running ahead of the CME flanks are real, 
our result provides the first observational evidence of such waves.
As such, global EUV waves can no longer be considered as single-pulse disturbances
as previous low-resolution observations 	
indicated \citep{Wills-Davey.EIT-wave-soliton.2007ApJ...664..556W}.		


As the EUV wave propagates outward in all directions, the ambient coronal loops 
are sequentially displaced at increasing distances, with delays in agreement with wave travel times.
These loops generally overshoot and start fast kink mode oscillations,
again, driven by a Lorentz restoring force produced by the displaced magnetic field.
These oscillations, signaling the {\it first} response to the arrival of the traveling disturbance, 
are another manifestation of its fast-mode wave nature 
that is the propagation of a perturbation 	
at the {\it highest} velocity supported by the medium. 
These oscillating loops do not participate in the CME expulsion,
but are only passively perturbed, while maintaining their topological integrity 
largely unchanged. They are thus not the footprint of a CME 	
\citep[cf.,][]{	
Attrill.EIT-wave-CME-Footprint.2007ApJ...656L.101A}.




Upon arrival at the coronal cavity, the global wave front and subsequent quasi-periodic wave
train continue to travel into it and exit from its far side.
Their velocities within the cavity being 50\% higher than outside
indicate higher \Alfven  	
and fast-mode speeds, 	
expected from the lower plasma density and higher magnetic field strength 
in a flux-rope cavity \citep{GibsonS.cavity.3D.2010ApJ...724.1133G, Berger.hot-bubble.2011Natur.472..197B}.
In other words, the same disturbance travels into the cavity
at velocities modified according to the local fast-mode speed.
(If the wave were to bypass the cavity in front of or behind it in projection, 
one would not expect such a velocity change.)
This also implies that this disturbance 	
must have penetrated through a topological separatrix surface that separates the
flux-rope, as a distinct flux system, from the surrounding corona. 
This contradicts the prediction that a CME-caused apparent wave cannot travel across such a surface 
\citep{ChenFP.EIT-wave-MHD.2002ApJ...572L..99C, Delannee.EITwv.stationary.2007A&A...465..603D}.
The EUV wave does, however, stop at the polar coronal hole boundary, another
separatrix surface, as observed before \citep{ThompsonB.EIT-wave-catalog.2009ApJS..183..225T}. 
This is likely because of the strong reflection due to the
large gradient of the wave speeds there \citep{Schmidt.Ofman.3D-MHD-eitwv.2010ApJ...713.1008S}. 
The lack of reflection signature at the coronal cavity may suggest a relatively small wave speed
gradient and thus a large transmission (versus reflection) coefficient, which requires
further investigation.

The QFP wave trains traveling 	
along the coronal funnel likely originate from the flare, 
as previously found \citep{LiuW.FastWave.2011ApJ...736L..13L}, 
but with some different frequency distributions. They are spatially
confined {\it within} the CME bubble, which may have produced a favorable condition
for their generation and/or detection. Their sudden termination at the CME front is likely 
due to damping generated by the discontinuity there that rapidly dissipates these waves,
and/or due to increased dispersion associated with the decrease of the fast-mode speed with height, 
as predicted in 3D MHD simulations \citep{Ofman.Liu.fast-wave.2011ApJ...740L..33O}.



\subsection{Discussion}
\label{subsect_conclude:discuss}



The forward inclination of the early EUV wave front
in the low corona has important implications \citep[e.g.,][]{GrechnevV.shock-EUV-wave-I.2011SoPh..273..433G}.
An observational consequence is a delayed arrival by, e.g., tens of seconds to minutes,
of its chromospheric counterpart, a Moreton wave 	
as indicated in recent preliminary results 	
\citep{WhiteS.2011Feb.Moreton.AIA.waves.2011SPD....42.1307W}.
While the early-stage EUV wave front is primarily shaped by its driver, 
as the wave travels further away, the propagation effect
will become increasingly influential.
If the EUV wave is a true MHD wave, refraction can deflect its ray trajectory 
away from regions of high wave speeds toward regions of low wave speeds		
\citep{Uchida.Moreton-wave-sweeping-skirt.1968SoPh....4...30U, 
Afanasyev.Uralov.shock-EUV-wave-I.2011SoPh..273..479A}.
Therefore, the shape and temporal evolution of the EUV wave front
has a seismological potential \citep{Ballai.global-corona-seism.2007SoPh..246..177B} 
to infer the spatial distribution of the fast-mode and \Alfven speeds 
and thus the coronal magnetic field strength.


The common periodicities of the quasi-periodic waves (especially those ahead of the CME)
and flare pulsations and the consistent extrapolated velocities
(see \sect{sect_QFP:FFT}) point to the possibility that
a cyclic disturbance, likely a fast-mode MHD wave, originates from the flare site	
\citep[cf.,][]{ChenFP.EIT-wave-not-flare-driven.2006ApJ...641L.153C}
and modulate the EUV waves both {\it ahead of} and {\it behind} the CME. 
The different physical environments
in which these waves propagate may further modify 	
their relative amplitudes by, e.g., a combination of 
phase velocity variations and wave dispersion,
giving rise to their different observed frequency distributions.
We also note that no clear wave signatures can be identified within 
$\sim$$150 \arcsec$ ($100 \arcsec$) from the eruption center
for the waves ahead of (behind) the CME front.
We speculate this being caused by 	
the line-of-sight integration of other more dominant features
near the eruption and/or disturbance amplitudes below the instrument sensitivity 
in the early stage.
Another possibility for such common periodicities 
is fast kink mode oscillations of loops in the peripheral
of the active region instigated by the CME eruption. In this case, the period 
$P = 2 l /(N c_k) $ for the fundamental mode ($N=1$)
gives the loop length
$l = P c_k/2 = (2 \min) \times (1 \Mm \ps) /2  \approx 60 \Mm$,
which is reasonable for large active region loops and comparable to the height range 
($\sim$110~Mm, \fig{tslice_QFP.eps}(i)) of the low-corona wave trains.
However, despite our best efforts, we could not identify such loop oscillations in AIA data.
Alternatively, if these oscillations are not free, linear kink modes, 
but are driven or nonlinear, their periodicities would be determined 
by the driver or nonlinearity, not by the loop length. 
These aspects require further investigation with 3D MHD models.


Recent observations \citep{HarraSterling.fast.leading.edge.EIT-wave.2003ApJ...587..429H, Zhukov.EIT-wave2004A&A...427..705Z, 
Kienreich.2009ApJ...703L.118K, Patsourakos.EUVI-fast-mode-wave.2009ApJ...700L.182P, 
ChenFP.Wu.Coexis-fast-mode.2011ApJ...732L..20C,
ChengX.ZhangJie.EUV-wave-CME-decouple.2012ApJ...745L...5C}
and simulations \citep{ChenFP.EIT-wave-MHD.2002ApJ...572L..99C, ChenFP.EIT-wave.2005ApJ...622.1202C,
CohenO.EITwave.non-wave.both.2009ApJ...705..587C, DownsC.MHD.EUVIwave.2011ApJ...728....2D,
DownsC.MHD.2010-06-13-AIA-wave.2012ApJ}
have led to a converged view about the multifaceted nature of global EUV waves.
That is, a {\it primary (outer) fast-mode MHD wave/shock}, sometimes associated with an 
\Ha Moreton wave \citep{ThompsonB.EIT-Moreton-wave.2000SoPh..193..161T,
AsaiA.2010AugX6.9.Moreton.AIA.wave.2012ApJ...745L..18A},
travels generally ahead of a {\it secondary (inner) slow disturbance} related to the CME-caused coronal restructuring.
Along the lines of recent suggestions \citep[e.g.,][]{DownsC.MHD.EUVIwave.2011ApJ...728....2D, 
Schrijver.Aulanier.2011Feb15.X2.AIAwv.2011ApJ...738..167S},
we propose that this wave/non-wave hybrid mechanism is a general rule, rather than an exception.
Under specific physical and observational circumstances, 
this general underlying mechanism may have different manifestations,
depending on which component has the dominant observational signature.
Taking the EUV wave reported by \citet{LiuW.AIA-1st-EITwave.2010ApJ...723L..53L}
as an example, the low-cadence, low sensitivity \sohoA/EIT could miss the 
leading faint, diffuse wave component of $240 \kmps$ 
and only capture the trailing strong, sharp non-wave fronts as slow as $80 \kmps$. 
This could cause a bias in the broad velocity distribution 
of 50--$700 \kmps$ obtained from EIT \citep{ThompsonB.EIT-wave-catalog.2009ApJS..183..225T}, 
the lower end of which is		
incompatible with fast-mode waves.
Likewise, if an expanding CME 	
drives a fast-mode wave or shock, depending on whether the stand-off distance
between the wave and CME front can be spatially resolved by the instrument, 
one could detect both or only one component.
When a separatrix surface lies in the way, the CME-caused slow component could stop,
while the fast-mode component would proceed forward.	
As such,	
the single term ``EIT waves" may have historically
included a variety of physically distinct phenomena 
\citep{Warmuth.Mann.3class.EIT.wave.2011A&A...532A.151W}. 



To date, \sdoA/AIA has observed a variety of EUV waves, including
those generated by mini-CMEs, emerging flux
\citep{ZhengRS.AIA.EUVwv.miniCME.2011ApJ...739L..39Z, ZhengRS.EUV-wave-EFR.2012ApJ...747...67Z},
and ``coronal cyclones" \citep{ZhangJun.LiuYang.AIA.cyclone.2011ApJ...741L...7Z}, 
as well as Kelvin-Helmholtz instability waves produced by velocity shear at CME flanks
(see \citealt{Ofman.Thompson.AIA.KH.instab.2010AGUFMSH14A..02O, Ofman.Thompson.AIA.KH.instab.2011ApJ...734L..11O}
for the first AIA EUV wave event reported by \citealt{LiuW.AIA-1st-EITwave.2010ApJ...723L..53L};
cf., \citealt{Foullon.AIA.KH.instab.2011ApJ...729L...8F} for a later event).	
These new observations, together with those presented here,
add to the multitude of the EUV wave phenomenon in specific and CME/flare eruptions in general.
A latest study of a large sample of $\geq$100 global EUV waves
observed by both \sdoA/AIA and \stereoA/EUVI has revealed an average
wave velocity of $700 \kmps$ (N.~Nitta et al.~2012, in preparation).
This is a strong indication that the primary EUV disturbance is a fast-mode MHD wave,
a conclusion supported by many recent studies 
\citep[e.g.][]{	
AsaiA.2010AugX6.9.Moreton.AIA.wave.2012ApJ...745L..18A}
and particularly the quasi-periodic wave trains ahead of the CME discovered here.
The origin of their periodicity will be a subject of future investigation.



\acknowledgments
{This work is supported by NASA \sdoA/AIA contract NNG04EA00C to LMSAL, 
WL and LO by NASA grant NNX11AO68G, and LO by NASA grant NNX09AG10G.
We thank Cooper Downs, Spiros Patsourakos, P.~F. Chen, and Keiji Hayashi for useful discussions
and the referee for constructive comments that helped improve this paper.
The \sohoA/LASCO CME catalog is generated and maintained at the CDAW Data Center by NASA 
and The Catholic University of America in cooperation with the Naval Research Laboratory. 
} 







\section*{Appendix A \\
  Technical Details of Data Analysis}
\label{sect_analysis}

As shown in \figs{mosaic_overview.eps} and \ref{tslice_fast.eps}, we selected
four complementing sets of narrow cuts (slits) for space--time analysis:
  (1) 15 radial cuts R$i \,\, (i=0,...,14)$ of $10\arcsec$ wide 
starting from the eruption center ($h=49 \arcsec$ above the limb at N$20.6\degree$),
which serve to track the early expansion of the CME and the generation of the EUV wave;
  (2) four azimuthal cuts A$i$		
that are $20\arcsec$--$30\arcsec$ wide with centers located at heights of 
$h=38 \arcsec, \, 88 \arcsec, \, 168  \arcsec, \, 258 \arcsec$ 			
and serve to track propagation at constant heights above the limb;
  (3) 12 vertical cuts V$i$	
of $20\arcsec$ wide starting at the limb, which serve to provide height dependent arrival times of
propagating disturbances, with cut V6 located at the eruption center,
cut V9 within the coronal cavity,	
and cut V11 at the coronal hole boundary;
and (4) three curved cuts C$i$	
 of $20\arcsec$ wide		
along the funnel in which the QFP waves propagate.
%
 \begin{figure*}[thbp]      
 \epsscale{0.9}
 \plotone{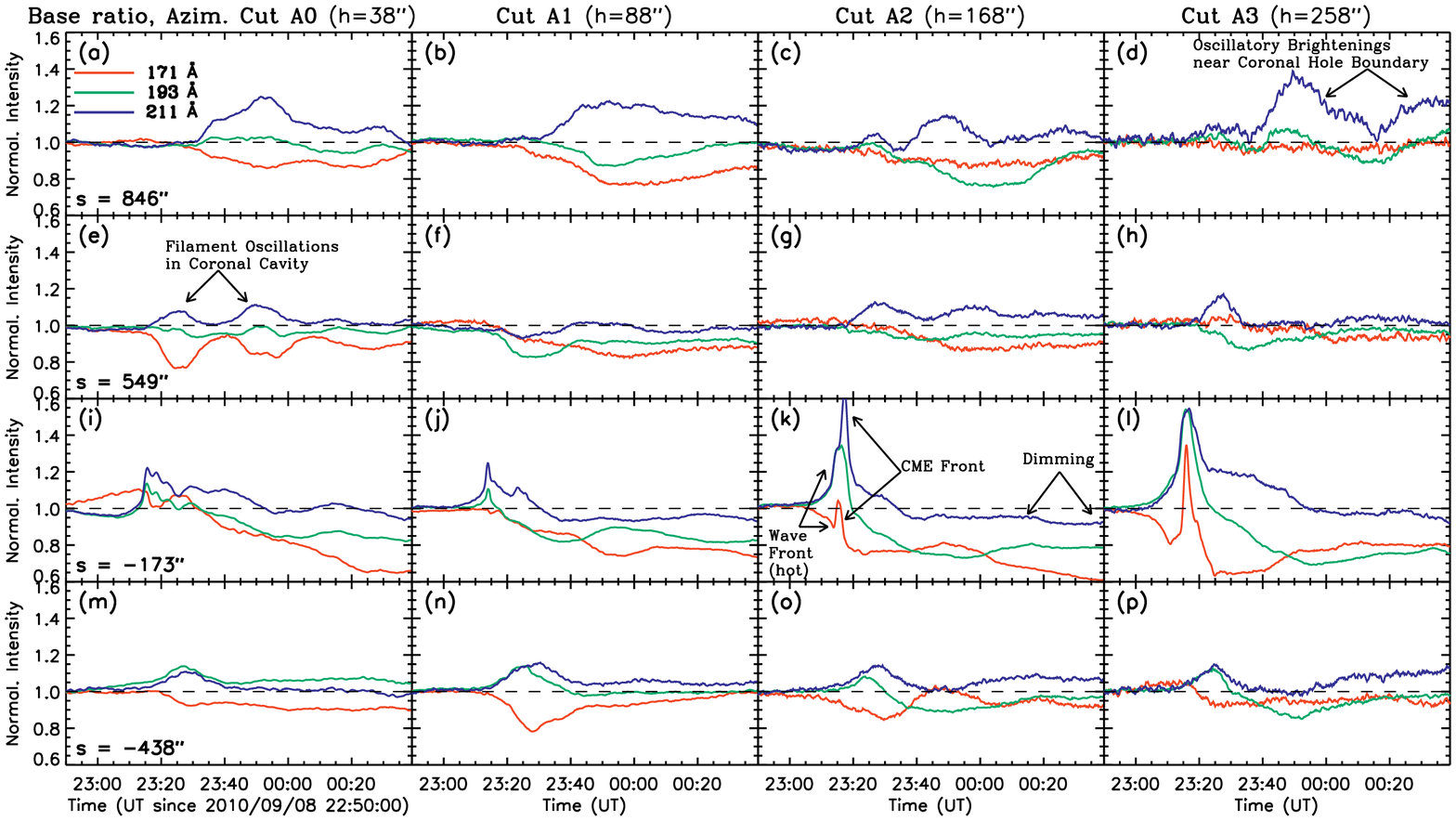}		
 \caption[]{
 Base ratio temporal profiles of emission intensity from azimuthal cuts
 at selected positions shown in Figure~\ref{tslice_overview.eps} (plus signs there).
 The general trend of darkening at 171~\AA\ and brightening at 193 and 211~\AA\
 indicates heating in the EUV wave pulse ahead of the CME.
 } \label{time_prof.eps}
 \end{figure*}
 \begin{figure*}[bht]      
 \epsscale{1.1}
 \plotone{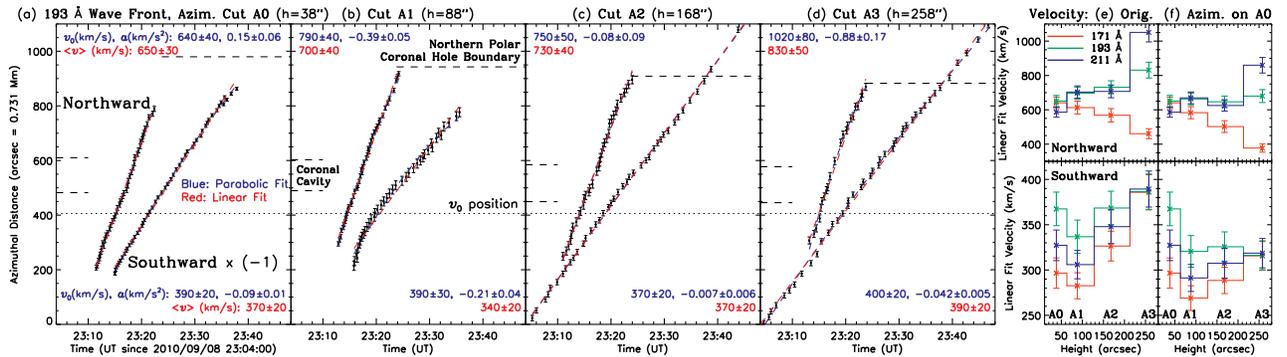}		
 \caption[]{	
 (a)--(d) Positions of the north- and southward 193~\AA\ EUV wave fronts
 obtained from azimuthal space-time diagrams as in Figure~\ref{tslice_overview.eps}.
 The sign of the southward distance is reversed.
 Overlaid dashed lines are parabolic (blue) and linear (red) fits, 
 labeled with fitted parameters. 
 The horizontal dotted line denotes the position where the velocity $v_0$ from parabolic fits
 is evaluated.
  (e) Height distribution of velocities from linear fits	
 to wave fronts on different azimuthal cuts at 171, 193, and 211~\AA.
  (f) Same as (e) but for velocities scaled onto the lowest cut A0, equivalent to angular velocities
 in the azimuthal direction with respect to the disk center.
 } \label{multi_vfit.eps}
 \end{figure*}
%

Each space--time diagram was obtained by composing
one-dimensional image profiles at different times along a cut.
To achieve desired color scaling suitable for different features,
we further applied base or running difference 	
by subtracting an initial or immediately previous image profile, 
or likewise, applied base or running ratio, 	
equivalent to differencing logarithmic profiles.
We used mouse-clicks to track a moving/propagating feature whose positional uncertainties,
depending on its sharpness, are typically no more than a few to $\sim$$10\arcsec$. 

Bear in mind that velocities measured this way are
projected velocities on the plane of the sky captured within the cut. They can deviate from 
the true plane-of-sky velocities by a factor of $1/\cos{\theta}$, if the cut is at an angle $\theta$ 
from the direction of propagation normal to an intensity front.
Significant deviations, however, only occur for large angle mis-alignments,
say, a factor of 2 overestimate for $\theta=60 \degree$.
Because this event occurs on the limb, we expect that such projection effects
are rather small. The selection of complementing cuts above
can also help minimize this potential bias.

\stereo A, $82\degree$ ahead of the Earth, 
provides complementary information at a 2.5~min cadence
of the CME-EUV wave decoupling, consistent with our 	
and earlier results 	
\citep[e.g.,][]{	
Patsourakos.EUVI-wave.2009SoPh..259...49P}.	
We thus chose 	
to focus on new AIA observations here, but
future studies using higher cadence \stereo data may help
constrain the 3D geometry of quasi-periodic wave trains.
	%

\section*{Appendix B \\
  General Properties of Global EUV Wave}
\label{append_sect_wave:gen}

\section*{	
  B1. 
Temperature Dependence: Heating and Cooling}
\label{sect_wave:T}

The general trend of darkening at 171~\AA\ and brightening at 193 and 211~\AA\
at the EUV wave front, followed by a recovery in the opposite direction, as shown in \fig{tslice_overview.eps},
indicates heating and subsequent cooling. 	
Qualitatively similar behaviors were observed by
\trace and \stereoA/EUVI \citep{Wills-DaveyThompson.TRACE-EUV-wave.1999SoPh..190..467W, 
DaiY.EUVI-wave.non-wave.2010ApJ...708..913D} and \sdoA/AIA
\citep{LiuW.AIA-1st-EITwave.2010ApJ...723L..53L, LongD.AIA.EUV-wave.2011ApJ...741L..21L,
Schrijver.Aulanier.2011Feb15.X2.AIAwv.2011ApJ...738..167S}.

We further take horizontal slices at selected positions from \fig{tslice_overview.eps} (plus signs there)
and show the resulting intensity profiles in \fig{time_prof.eps}.
The large variations near the eruption (panels~(k) and (l))
are due to the erupting material 	
and subsequent dimming, while the rest, moderate variations result from the global EUV wave.
In particular, the CME front in panel~(k) appears as a sharp enhancement in
all three channels, suggesting an expected density increase. 
Immediately ahead of it is a dip at 171~\AA\ and a shoulder 
at 193 and 211~\AA, indicating a heated sheath surrounding the CME front.

Within the EUV wave pulse (e.g., panel~(n)), it usually takes several to 20 minutes to reach the 	
maximum intensity variation at each wavelength, followed by a comparable or longer		
recovery lasting up to tens of minutes.
The maximum 193~\AA\ enhancement is about 5--10\% 
above the pre-event level, while the maximum 211~\AA\ enhancement and 171~\AA\ reduction
are about two times larger on the 10--20\% level and delayed by a few minutes
from the 193~\AA\ maximum \citep{LiuW.AIA-1st-EITwave.2010ApJ...723L..53L, MaSL.AIA.shock.2011ApJ...738..160M}.
This indicates prompt plasma heating within minutes across the characteristic temperatures of the
171, 193, and 211~\AA\ channels, from $0.8 \MK$ to $1.6 \MK$ and then $2.0 \MK$, respectively
(\citealt{ODwyer.AIA-T-response.2010A&A...521A..21O, Paul.Boerner.initial.AIA.calib.2011SoPh..tmp..316B}; cf., \citealt{Kozarev.AIA.EUVwv.2010Jun12.2011ApJ...733L..25K}).
The subsequent recovery after the maximum toward the pre-event level is a cooling phase.
Such emission variations were largely reproduced in MHD simulations
and ascribed to heating by initial compression followed by cooling by restoring rarefaction
in an adiabatic process associated with a compressible wave \citep{DownsC.MHD.2010-06-13-AIA-wave.2012ApJ}.


This general trend is evident in the quiet Sun to the south of the eruption.
To the north, it is complicated by organized structures, 
such as the coronal cavity/filament and polar coronal hole boundary
producing oscillatory intensity variations (panels~(d) and (e)).

\section*{B2.	
Kinematics and General Deceleration}
\label{sect_wave:azimuth-kinem}

To measure the velocity of the global EUV wave front, we visually identified the initial intensity
change, regardless of brightening or darkening, at each distance in base-ratio space--time diagrams 
(e.g., \fig{tslice_overview.eps}) obtained from azimuthal cuts shown in \fig{mosaic_overview.eps}(b).
We have also tested automated methods 
\citep[e.g., see][]{LiuW.AIA-1st-EITwave.2010ApJ...723L..53L, LongD.STEREO.EUV-wave.Dispersion.2011A&A...531A..42L},
but all failed to consistently detect the wave front because of the complex intensity variations, 
as shown in \fig{time_prof.eps}.	
Yet, the visual detection,	
with acceptable uncertainties, yielded the most consistent result.
	
The resulting EUV wave front positions versus time are shown in \fig{multi_vfit.eps}	
for the 193~\AA\ channel, as an example, and fitted with  
linear (red dashed) and parabolic (blue dashed) functions. 
In general, both functions fit the data nearly equally well within uncertainties, 
but parabolic fits yield slightly smaller $\chi^2$s and indicate 
decelerations up to	
to $-0.88 \pm 0.17 \kmpss$
(except for the acceleration 	
in panel~(a)).
At the lowest cut A0 (height $h=38 \arcsec$), linear fits give
an average velocity of $650 \pm 30 \kmps$ for the northward wave 
and $\sim$50\% slower at $370 \pm 20 \kmps$ for the southward wave. 
This north-south asymmetry generally holds at different heights.
There is a weak increase of the velocity with height, 	
but not significant beyond uncertainties.

The EUV wave velocities measured at different wavelengths 
show substantial scatter (\fig{multi_vfit.eps}(e) and (f)),	
consistent with those found by \citet{LongD.EUVI-wave.2008ApJ...680L..81L, LongD.AIA.EUV-wave.2011ApJ...741L..21L}, 
but their identified trend of higher velocities in cooler lines is not present here.
At a given height, the velocity generally decreases from 193 to 211 and 171~\AA,
likely due to the delay of the same trend noted in Appendix~B1.	
 \begin{figure}[thbp]      
 \epsscale{0.9}
 \plotone{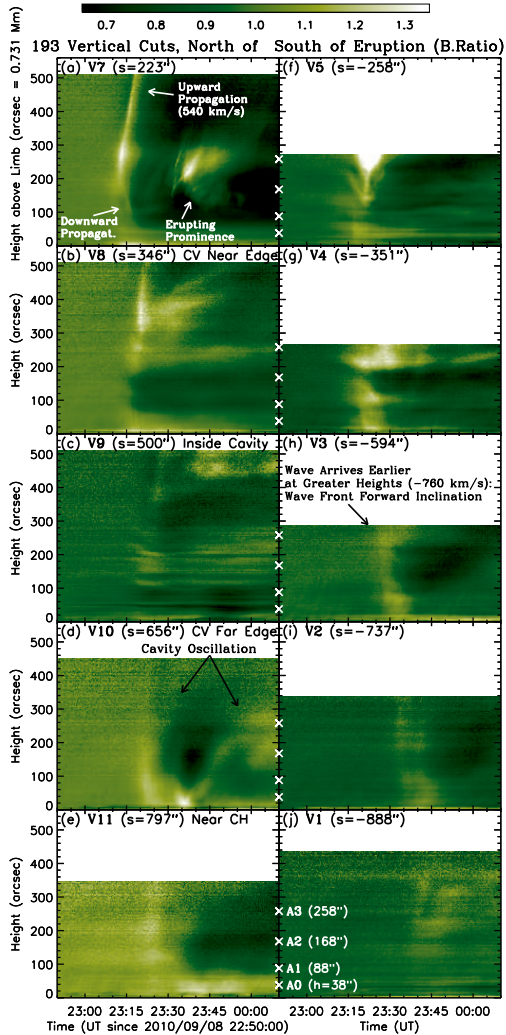}		
 \caption[]{
 193~\AA\ base ratio space--time diagrams from selected vertical cuts 	
 in Figure~\ref{mosaic_overview.eps}(f),
 showing height-dependent arrival time of the EUV disturbance.
 Each panel is labeled with the azimuthal position $s$ of the vertical cut,
 as marked by the cross signs in \fig{tslice_overview.eps}(c).
 The cross signs here on the $y$-axis	
 mark the heights of the azimuthal cuts A0--A3.
 } \label{tslice_vcut.eps}
 \end{figure}
 \begin{figure}[thbp]      
 \epsscale{0.9}
 \plotone{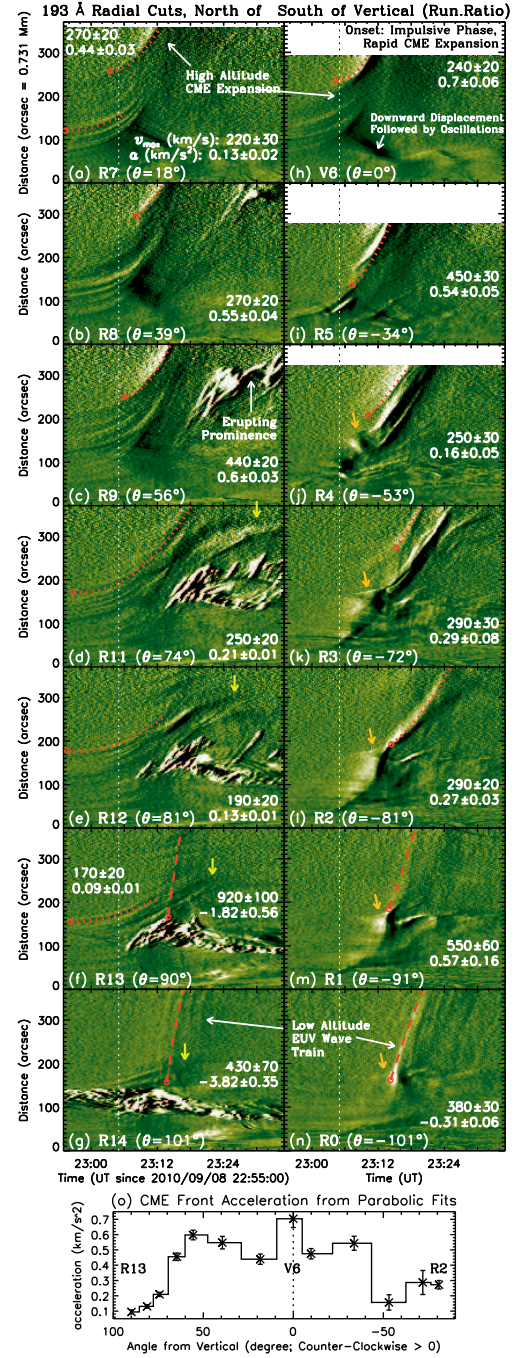}		
 \caption[]{
   (a)--(n) Running ratio space--time diagrams at 193~\AA\ from selected radial cuts R$i$
 and vertical cut V6 shown in \figs{mosaic_overview.eps}(a) and (f).
 Distance is measured from the eruption center		
 located at a height of $49 \arcsec$ on V6.
 Note the acceleration of the CME front and followup loops at high altitudes
 and the generation of the EUV wave at low altitudes.
 The red dotted and dashed lines are parabolic fits to the CME and EUV wave fronts, respectively, 
 labeled with the fitted acceleration (in $\kmpss$) and maximum velocity 
 (in $\kmps$, i.e., the initial velocity for deceleration
 or the final velocity at $s=370 \arcsec$, top distance of the plot, for acceleration).
 The vertical white dotted line marks the onset of the rapid CME expansion
 and the flare impulsive phase at 23:05~UT.
   (o) The bottom panel shows the angular distribution of the CME acceleration 
 measured from these cuts.	
 The angle is measured from the local vertical in the counter-clockwise direction.
 } \label{tslice_radial.eps}
 \end{figure}
%

\section*{	
  B3. 
Height Dependence and Forwardly Inclined Wave Fronts}		
\label{sect_wave:height-vcut}

The height dependence of the EUV wave can be examined through its arrival 
at vertical cuts defined in \fig{mosaic_overview.eps}(f).	
As shown in the resulting space-time diagrams of \fig{tslice_vcut.eps},
	%
not far from the eruption (top three rows), 
a general pattern, similar across different AIA channels, is as follows.
In the low corona ($h \lesssim 300 \arcsec = 220 \Mm$), the initial intensity variation
forms a stripe of a negative slope, indicating that the arrival of the disturbance 
at the vertical cut first occurs at a high altitude and then progresses toward lower altitudes;
in the high corona ($h \gtrsim 300 \arcsec$), the trend is opposite.
For example, over a height range of $200\arcsec$, the downward progression 	
at cut~V3 (south of the eruption)	
results in a 3~min delay, translating to an apparent vertical velocity of $-760 \kmps$, 
while the upward progression at cut~V7 (north of the eruption) 		
produces a 4.5~min delay and an apparent velocity of $540 \kmps$.
This pattern means that the leading front of the EUV wave near the eruption has a sphere-like shape
\citep{Veronig.dome-wave.2010ApJ...716L..57V}:
viewed from the eruption center, the front is upwardly concave in the high corona and 
{\it forwardly inclined} toward the solar surface in the low corona,
as further discussed in \sect{sect_conclude}.

The delayed arrival at the vertical cut V8	
shown in \fig{tslice_QFP.eps}(k), for example, can be used to reconstruct the 
corresponding wave front. 	
For each height $h$, we first interpolated the azimuthal wave velocity $v(h)$
from those average velocities measured on cuts A0--A3 at different heights (see \fig{multi_vfit.eps}). 
We then obtained the azimuthal distance traveled from the actual arrival time $t(h)$ 
to a reference arrival time of $t(0)=$23:16:55~UT at the coronal base ($h=0$),
$ \Delta s(h) = [ t(0) - t(h) ] v(h) $.
This distance was added to the azimuthal position $s_{\rm cut}(h)$ of this vertical cut
measured from the eruption center,
to obtain the position of the instantaneous wave front at $t(0)$,
$ s(h) = s_{\rm cut}(h) + \Delta s(h)$.
Repeating this for all heights and then transforming the azimuthal positions back to the Cartesian coordinates
gave the resulting wave front, as shown in red in \fig{tslice_QFP.eps}(l).


\section*{Appendix C \\
  CME Kinematics and Early Expansion}	
\label{sect_decouple:CME-append}

\fig{tslice_radial.eps} shows the expansion in different directions
of coronal loops involved in the CME, as captured in radial cuts 	
defined in \fig{mosaic_overview.eps}(a). 
Strong acceleration ($a > 0.4 \kmpss$) only occurs within an angular range 
$-50 \degree \leq \theta \leq 70 \degree$ of the local vertical 
at the eruption center (panel~(o)). In this range, the CME experiences 
a two-phase, slow to fast expansion \citep{SterlingMoore.slow-fast.2005ApJ...630.1148S}, 
the transition of which occurs at 23:05~UT, the beginning of the flare impulsive phase.
The early-phase has an acceleration $\lesssim$$0.1 \kmpss$ 
and maximum velocity $\sim$$100 \kmps$ (panel~(a)).
The largest acceleration of $0.70 \pm 0.06 \kmpss$ in the late-phase takes place in the vertical direction
(panel~(h)), reaching $240 \pm 20 \kmps$ at the edge of the AIA FOV.
Comparable acceleration possibly continues into the 
\sohoA/LASCO FOV (1.5--30~$R_\sun$), as shown by the dotted line in the inset 
of \fig{flare_lc.eps}(c). The CME then decelerates at $a=-0.019 \pm 0.005 \kmpss$
with an initial and average velocities of $v_{\rm max}=990 \pm 70 \kmps$			
and $\langle v \rangle = 820 \pm 40 \kmps$,
according to the CME catalog (https://cdaw.gsfc.nasa.gov/CME$\_$list).

Beyond the angular range of $-50 \degree \leq \theta \leq 70 \degree$,
the CME acceleration remains below $0.3 \kmpss$, with no two-phase distinction
but a north-south asymmetry.
  To the north of the vertical, the expanding loops gradually slow down and/or fade away
(panel~(e)). This first occurs at the lowest cut and progresses toward cuts 
at greater heights, as indicated by the short yellow arrows.
  To the south, the early slow expansion is absent,
but a sudden brightening	
caused by a rapid {\it downward} expansion of some loops (see Movie~1C),
indicated by the short orange arrows, takes place sequentially from cut R4 to R0. 
It is at the outermost edge of this brightening near the coronal base,
$s_0= 150 \arcsec = 110 \Mm$ from the eruption center,
that the southward global EUV wave is launched at 23:14:20~UT (panels~(m) and (n)). 
Its northward counterpart is first detected at a similar distance 3~min earlier.

We note in panel~(h) that, in the vertical direction,
some loops at the eruption center are pushed downward 
from an initial height of $h_0= 150 \arcsec$
(with the $49 \arcsec$ height of the origin of the cut included), followed by
transverse oscillations.
In lateral directions, at the onset of the rapid CME expansion (vertical dotted line),	
all participating loops are also located approximately 
beyond a distance $s_0 = 110 \Mm$ from the eruption center.
This indicates that the rapid expansion involves 	
the entire active region of $r_{\rm AR}= 110 \Mm$ in radius and originates 
at its center of height $h_0=110 \Mm$. 	
These observations imply that this expansion,
with a significant downward component, may play a key role in driving the EUV wave
(see more discussion in \sect{subsect_conclude:proposal}).

\section*{Appendix D \\
  Estimates of EUV Wave Energy and Cavity Magnetic Field}
\label{append:energy-seism}

We can utilize the oscillations of the coronal cavity to estimate the energy budget
of the global EUV wave \citep[see, e.g.,][]{Ballai.EIT-wave-trigger-loop-oscil.2005ApJ...633L.145B}
and the cavity magnetic field strength, i.e., by coronal seismology.
These are order-of-magnitude estimates with comparable uncertainties.

The cavity has a diameter $D_{\rm cav} = 135 \arcsec = 99 \Mm$ measured on azimuthal cut~A2
in \fig{tslice_QFP.eps}(c). Its embedded filament, likely underneath the axis
of the flux rope running along its full length
\citep[e.g.,][]{LowBC.Hundhausen.flux.rope.prominence.1995ApJ...443..818L}, 
has been quite stable for a week before this event.	
It was best seen a few days earlier (say, September 4, at http://www.solarmonitor.org),
located at N$50 \degree$ on the disk, spanning a longitude range of $35\degree$ 
and a length of $L_{\rm cav} =  273 \Mm$.		
Assuming the cavity being a cylinder of such a dimension,	
its volume is $V_{\rm cav}= \pi (D_{\rm cav}/2)^2 L_{\rm cav}$. 
We adopt an average electron density of $n_e= 1\E{9} \pcmc$ of the tenuous cavity
and its dense filament combined
\citep{VasquezA.cavity.T.n.2009SoPh..256...73V}
and thus $\rho = n_e m_p = 1.7 \E{-15} \g \pcmc$. Its total mass
is $m_{\rm cav} = \rho V_{\rm cav} = 3.5 \E{15} \g$, comparable to a typical CME mass
\citep{Aschwanden.STEREO.dimming.CME.mass.2009ApJ...706..376A}.
Now, take a velocity amplitude of $v_{\rm osc}= 9 \kmps$, 
we obtain the kinetic energy of the cavity of
$E_{k, \rm cav}= m_{\rm cav} v_{\rm osc}^2 /2 \sim 2\E{27} \ergs $
and energy per unit length along its axis of $e_{k, \rm cav} = E_{k, \rm cav} / L_{\rm cav}$.
We further assume such an energy is uniformly distributed in space at the same projected distance
from the eruption center, $s = 450 \arcsec = 329 \Mm$, measured on cut~A2 (\fig{tslice_QFP.eps}(c)),
as a consequence of a surface-like wave in the low corona.
Then, the global EUV wave energy is 
$E_{\rm wv}= e_{k, \rm cav} (2 \pi s) \sim 1 \E{28} \ergs$.
Both energies are 3--4 orders of magnitude smaller
than the kinetic energy $E_{k, \rm CME} \sim 1 \E{31} \ergs$ of the CME,
assumed at the same mass as the cavity
moving at $820 \kmps$ measured by \sohoA/LASCO.
This wave energy is, however, 100 times greater than that of another
EUV wave and of a typical nanoflare \citep{Ballai.EIT-wave-trigger-loop-oscil.2005ApJ...633L.145B}.

These transverse oscillations are likely fast kink modes.
In this case, the period for the $N$th harmonics is $P = 2l/(N c_k)$, 
where $l$ is the loop length and $c_k \sim v_A \sim v_f$
is the kink speed in low-$\beta$ plasma with a loop density of
the same order of the surrounding plasma
($v_A$ and $v_f$ being the \Alfven and fast-mode speeds, respectively). 
Using the measured average period of 27.6~min for the fundamental mode $N=1$
and assuming $c_k = 980 \kmps $, we have $l = P c_k/2 \approx 800 \Mm
\approx 3 L_{\rm cav}$, where $L_{\rm cav} = 273 \Mm$ is the cavity's axial length
measured above. This gives a pitch angle of $\theta= \arccos(L_{\rm cav}/l) = \arccos (1/3) \sim 70 \degree $,
implying an expected, highly twisted flux rope \citep{CarterB.ErdelyiR.kink.oscil.twisted.tube.2008A&A...481..239C}.
If we adopt a typical density of $n_e= 2 \E{8} \pcmc$ 
\citep[$\rho= n_e m_p = 3 \E{-16} \g \pcmc $, from][]{SchmitDon.GibsonS.cavity-density-2e8.2011ApJ...733....1S}
for the coronal cavity, we obtain a magnetic field strength of
$ B = v_A \sqrt{4 \pi \rho} \sim 6 \G$.









\end{document}